\documentclass[
    prd, aps, twocolumn, nofootinbib, showpacs, superscriptaddress
]{revtex4-2}
\usepackage{graphicx}
\usepackage[utf8]{inputenc}
\usepackage{mathtools}
\usepackage{xcolor}
\usepackage{adjustbox}
\usepackage{placeins}
\usepackage[T1]{fontenc}
\usepackage{lipsum}
\usepackage{bm}
\usepackage{graphicx}
\usepackage{latexsym}
\usepackage{rotating}
\usepackage{natbib}
\usepackage{amsmath, amssymb, amsthm, amsfonts}
\usepackage[T1]{fontenc}
\usepackage{hyperref}
\usepackage{aas_macros}

\usepackage{soul}

\newcommand{\lambdatilde}{\ensuremath{\bar{\lambda}}}

\newcommand{\lambdazero}{\ensuremath{\bar{\lambda}^{(0)}_0}}
\newcommand{\lambdak}{\ensuremath{\bar{\lambda}^{(k)}_0}}

\newcommand{\mchirp}{\ensuremath{\mathcal{M}_c}}
\newcommand{\hubble}{\ensuremath{H_0}}
\newcommand{\dl}{\ensuremath{D_L}}
\newcommand{\xgw}{\ensuremath{d^{\text{GW}}}}

\newcommand{\msun}{\ensuremath{M_{\odot}}}

\newcommand{\mdet}{\ensuremath{m_{\text{det}}}}
\newcommand{\Ndet}{\ensuremath{N_{\text{det}}}}
\newcommand{\msource}{\ensuremath{m_{\text{source}}}}
\newcommand{\bomega}{\ensuremath{\mathbf{\Omega}}}
\newcommand{\lcdm}{\ensuremath{\Lambda\text{CDM}}}
\newcommand{\zngc}{\ensuremath{z_{\text{NGC4993}}}}
\newcommand{\hubbleunit}{\ensuremath{\text{km s}^{-1}/\text{Mpc}}}
\newcommand{\numeos}{{29}}
\newcommand{\bilby}{BILBY}
\newcommand{\pbilby}{PARALLEL\_BILBY}
\newcommand{\dynesty}{DYNESTY}

\usepackage[]{xcolor}
\definecolor{linkcolor}{rgb}{0.0,0.3,0.5}
\definecolor{darkgreen}{rgb}{0.0, 0.5, 0.0}
\definecolor{darkcyan}{rgb}{0.0, 0.5,0.5}

\begin{document}
\title{Cosmology with Love: 
Measuring the Hubble constant using neutron star universal relations
}

\author{Deep Chatterjee}
\affiliation{Center for AstroPhysical Surveys, National Center for Supercomputing Applications, Urbana, IL, 61801, USA}
\affiliation{Illinois Center for Advanced Studies of the Universe, Department of Physics, University of Illinois at Urbana-Champaign, Urbana, IL 61801, USA}

\author{Abhishek Hegade K. R.}
\affiliation{Illinois Center for Advanced Studies of the Universe, Department of Physics, University of Illinois at Urbana-Champaign, Urbana, IL 61801, USA}

\author{Gilbert Holder}
\affiliation{Illinois Center for Advanced Studies of the Universe, Department of Physics, University of Illinois at Urbana-Champaign, Urbana, IL 61801, USA}
\affiliation{Canadian Institute for Advanced Research, Toronto, Ontario M5G 1M1, Canada}

\author{Daniel E. Holz}
\affiliation{Department of Physics, Department of Astronomy and Astrophysics, Enrico Fermi Institute, and Kavli Institute for Cosmological Physics, University of Chicago, Chicago, Illinois 60637, USA}

\author{Scott Perkins}
\affiliation{Illinois Center for Advanced Studies of the Universe, Department of Physics, University of Illinois at Urbana-Champaign, Urbana, IL 61801, USA}

\author{Kent Yagi}
\affiliation{Department of Physics, University of Virginia, Charlottesville, VA 22904, USA.}

\author{Nicol\'as Yunes}
\affiliation{Illinois Center for Advanced Studies of the Universe, Department of Physics, University of Illinois at Urbana-Champaign, Urbana, IL 61801, USA}

\begin{abstract}
Gravitational-wave cosmology began in 2017 with the observation of the gravitational waves emitted in the merger of two neutron stars, and the coincident observation of the electromagnetic emission that followed. 
Although only a $30\%$ measurement of the Hubble constant was achieved, future observations may yield more precise measurements either through other coincident events or through cross correlation of gravitational-wave events with galaxy catalogs. 
Here, we implement a new way to measure the Hubble constant without an electromagnetic counterpart and through the use of the binary Love relations.
These relations govern the tidal deformabilities of neutron stars in an equation-of-state insensitive way. Importantly, the Love relations depend on the component masses of the binary in the source frame.  
Since the gravitational-wave phase and amplitude depend on the chirp mass in the observer (and hence redshifted) frame, one can in principle combine the binary Love relations with the gravitational-wave data to directly measure the redshift, and thereby infer the value of the Hubble constant.
We implement this approach in both real and synthetic data through a Bayesian parameter estimation study in a range of observing scenarios.
We find that for the LIGO/Virgo/KAGRA design sensitivity era, this method results in a similar measurement accuracy of the Hubble constant to those of current-day, dark-siren measurements. 
For third generation detectors, this accuracy improves to $\lesssim 10\%$ when combining measurements from binary neutron star events in the LIGO Voyager era, and to $\lesssim 2\%$ in the Cosmic Explorer era.

\end{abstract}


\maketitle

\section{Introduction} \label{sec:intro}

The inference of cosmological parameters like the Hubble constant, \hubble, using
gravitational waves (GWs) hinges on the \emph{standard siren}
approach~\citep{schutz_1986, holz_hughes_2005}. 
The central idea is to measure the luminosity distance, \dl, from the GW data while simultaneously
identifying an electromagnetic (EM) signal from the source.
The independent measurement of $\dl$
and the cosmological redshift, $z$, leads to a measurement of \hubble. In the absence
of a counterpart, clustering of {\hubble} measurements from potential host galaxies
for a large sample of events also leads to a \emph{statistical} measurement
of \hubble~\citep{2020PhRvD.101l2001G}. This idea found its first application in
the simultaneous panchromatic observations of GWs, gamma-rays, optical, and infrared radiation from
the binary neutron star (BNS) merger seen by the Advanced Laser Interferometer
Gravitational-wave Observatory (LIGO)~\cite{advanced_ligo} and Advanced Virgo~\cite{advanced_virgo}
detectors, GW170817~\citep{2017PhRvL.119p1101A}. The
identification of the host galaxy, NGC~4993, led to an independent 
measurement of {\hubble}~\citep{2017Natur.551...85A}. Also, the $\sim 16 {\text{deg}^2}$
sky-localization led to a statistical measurement of {\hubble}, agnostic of the host galaxy 
information~\citep{2019ApJ...871L..13F}. Such independent measurements are crucial
in the light of the recent tension in the value of {\hubble} measured from observations
of the early and late-time universe~\citep{2019NatAs...3..891V}.

A different approach of estimating the distance-redshift relation, solely using GWs
from merging BNS systems, was first proposed by \citeauthor{messenger_read_2012}~\cite{messenger_read_2012}.
Measuring the redshift is challenging because while the amplitude of the GW encodes information
about \dl, the mass parameters are degenerate with the redshift, resulting in the measurement of
the \emph{redshifted} mass at the detector as opposed to the true source-frame
mass, i.e.,~$\mdet = (1 + z)\msource$. 
However, matter effects in BNS inspirals, characterized by
the tidal deformability parameter, \lambdatilde, breaks this degeneracy since the tidal deformability is a function
of the source-frame mass, \msource. This feature has been exploited in the literature via an expansion of
\lambdatilde~in terms of \msource~\citep{damour_2012, delpozzo_2013, agathos_2015}. However,
the expansion coefficients are dependent on the NS equation of state (EoS), which is uncertain. This
implies that, in the absence of \textit{a priori} information about the EoS, the expansion coefficients
are free parameters in data analysis, and an extraction of the {\hubble} is difficult.

This degeneracy, however, can be broken through a set of universal relations discovered by \citet{yagi_yunes_2016, yagi_yunes_2017}~(hereafter YY17). 
Their work has shown that there exists tight relations between these expansion coefficients
that are insensitive to the EoS. 
Using these so called $\lambdazero - \lambdak$ \textit{binary Love} relations, knowing
\emph{one} of the expansion coefficient (which we call $\lambdazero$) determines the others, and
this reduces the number of free parameters drastically.
Since the {\lambdazero} is universal, the measurements
can be stacked using data from
gold-plated BNS events for which the source frame mass, or equivalently the redshift, is known.
Subsequently, its value may be fixed to express $\lambdatilde = \lambdatilde(\msource)$ for
future BNS detections leading to a measurement of $\msource$, or equivalently the redshift, and thus {\hubble}.

In this paper, we use the YY17 prescription to construct a $\lambdazero - \lambdak$ relation
using EoSs that satisfy the current constraint on mass and radii of NSs from LIGO/Virgo for GW170817~\cite{2018PhRvL.121p1101A}
and the same from by the Neutron Star Interior Composition Explorer (NICER) measurements of
the millisecond pulsar, PSR J0030+0451~\cite{2019ApJ...887L..24M}.
We then perform Bayesian parameter estimation on GW170817 data, employing this relation to measure
the free coefficient, $\lambdazero$, appearing in the expansion. We then analyze the
prospects of measuring {\hubble} by performing Bayesian parameter estimation on a set of
simulated (synthetic) BNS signals across different detector eras. While the {\hubble} measurement from
individual events may not be very constraining due to the distance-inclination degeneracy in
GW parameter estimation, we show that by combining the result from multiple events, the stacked
measurement of {\hubble} converges to the true value. We find that for the design sensitivity
LIGO/Virgo/KAGRA/India (HLVKI) era, the measurement accuracy is comparable to the current \emph{dark siren}
measurements~\cite{Fishbach_2019,Soares_Santos_2019}, or from recent counterpart measurement, assuming
the association of GW190521 and ZTF19abanrhr,~\cite{Abbott_2020,Graham_2020} shown
in Refs.~\cite{mukherjee2020measurement, 2020arXiv200914057C}. This accuracy improves to $\sim 10\%$ in the
Voyager era assuming $\mathcal{O}(10^2)$ BNS events, and to $\sim 2\%$ for $\mathcal{O}(10^3)$ detections
in the Cosmic Explorer (CE) era. Ultimately, the accuracy of the measurement of {\hubble} will be limited
by systematic uncertainties in the binary Love method, and we study what limits these systematic uncertainties
place on future measurements of {\hubble}.

The organization of this paper is as follows. In Sec.~\ref{sec:lambda_0}, we summarize the
universal binary love relations in brief. We show the construction of the $\lambdazero - \lambdak$
relation for this work. In Sec.~\ref{sec:pe}, we show our measurement for {\lambdazero} by performing
Bayesian parameter estimation on GW170817 data. In Sec.~\ref{sec:hubble}, we use the results of {\lambdazero}
and simulated BNS events across different generations of ground-based detectors to show the prospects
of measuring {\hubble} using multiple events. In Sec.~\ref{sec:uncertainties}, we consider the systematic
errors that could arise in the measurement of {\hubble} due to uncertainty in measurement of {\lambdazero},
and from uncertainties in  the {$\lambdazero - \lambdak$} relations. Finally, we conclude in
Sec.~\ref{sec:conclusion}. We follow the conventions of Ref.~\cite{Carroll:2004st}, and when necessary,
use geometric units $G = 1 = c$.

\section{Binary Love Relations}\label{sec:lambda_0}
The quasi-circular inspiral of a compact binary system is described under the
post-Newtonian (PN) formalism~\cite{Blanchet_1995}, where the system parameters, like
masses and spins, appear in the coefficients of the expansion at different PN orders. While the GW signal
from a BNS system has similar morphology to an analogous binary black hole (BBH) system
during the early inspiral phase, strong gravitational effects in the late inspiral stage deforms the NSs,
leading to additional multipolar deformations that enhances GW emission and leads to an earlier
merger~\cite{2008PhRvD..77b1502F}. The deformation of the NS is characterized by the electric-type, $\ell=2$, 
tidal deformability parameter, $\lambdatilde = (2k_2/3)C^{-5}$, where $C = M/R$
is the compactness of a NS with mass $M$ and radius $R$, which therefore depends on the EoS, and $k_2$ is the relativistic Love
number~\cite{Hinderer_2008}. This tidal deformation modifies the binding energy of the binary, which in
turn modifies the chirping rate, and therefore the waveform phase, with finite-size
effect appearing first at 5PN order ~\cite{2008PhRvD..77b1502F}. In a BNS system, the
waveform phase will have modifications that will depend on the tidal deformability of each
binary component, but there exist certain EoS-insensitive relations that relate them. We present
them here in brief, and refer the interested reader to YY17 for further details (see
also~\cite{De:2018uhw} for a similar EoS-insensitive relation between the tidal deformability
of each component). 
\begin{figure}[hp!]
    \centering
    \includegraphics[width=1.0\columnwidth, trim=0cm 0cm 0cm 0cm]{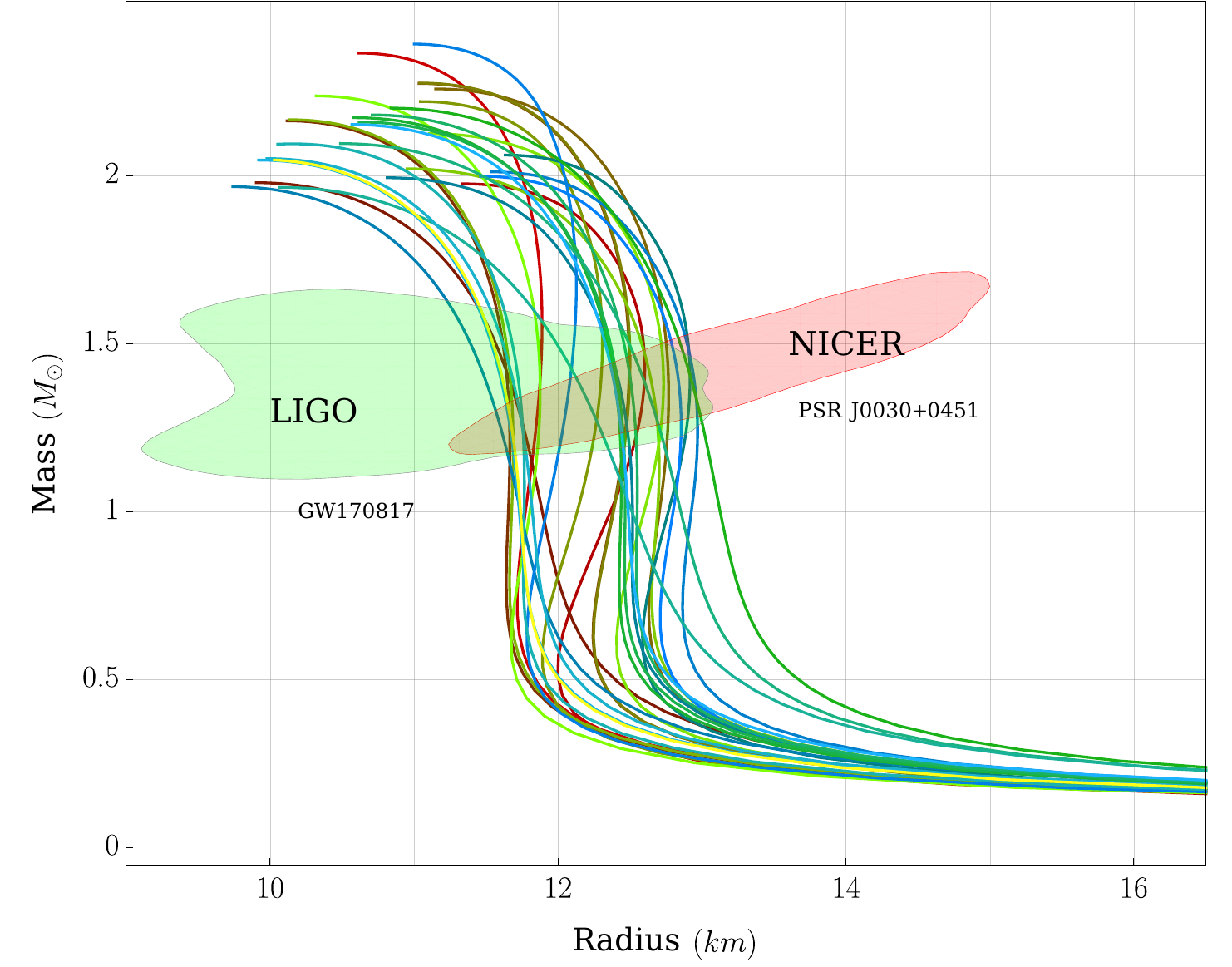}
    \\
    \includegraphics[width=1.0\columnwidth, trim=0.6cm 0cm 1.6cm 0cm]{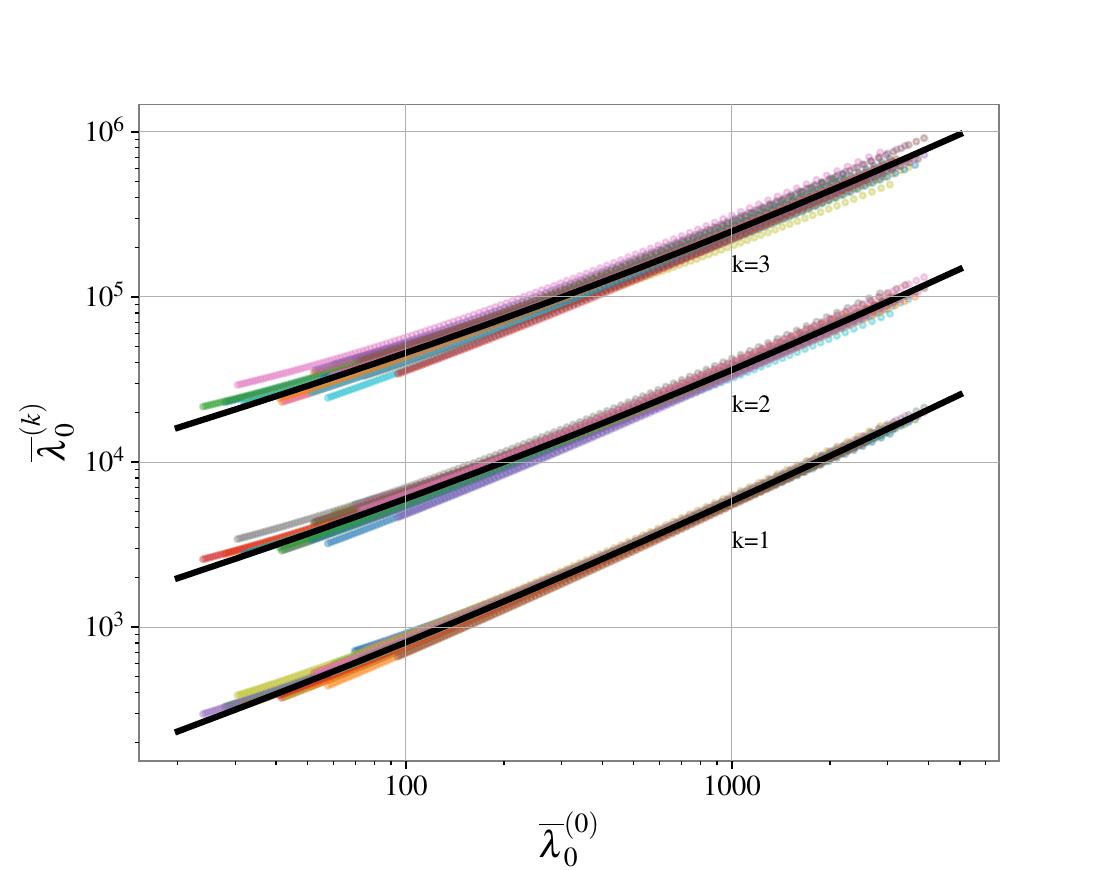}
    \\
    \includegraphics[width=1.0\columnwidth, trim=0.6cm 1cm 1.6cm 0cm]{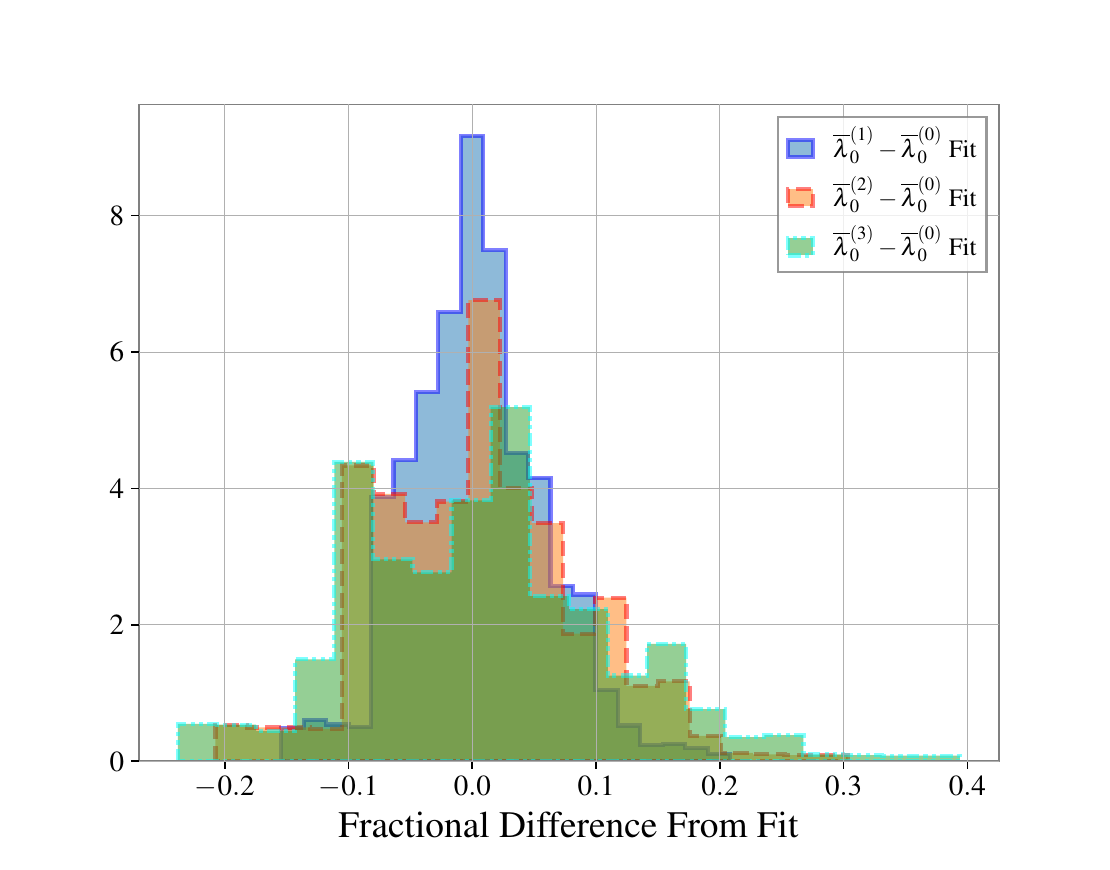}
    \caption{
    \textbf{Upper panel}: Mass-radius relations for
    {\numeos} piecewise polytropic EoSs that are consistent with the LIGO/Virgo
    and NICER observations to 68\% confidence.
    \textbf{Middle panel}: $\lambdak = \lambdak(\lambdazero)$ relations for the
    same set of EoSs, together with the fit in thick lines. Observe
    that the $\lambdazero -\lambdak$ relations are EoS insensitive and the fit
    recovers the average. 
    \textbf{Bottom panel}: Distribution of the fractional residual errors from the fit
    for each coefficient.
    }
    \label{fig:mass_radius_relations}
\end{figure}

While the internal composition of NSs is extremely complex, certain relations among
some observables, like the moment of inertia, the quadrupole moment, and the tidal
deformability, are insensitive to the details of the microphysics~\cite{Yagi_2013,Yagi_2013_PRD}
(see Refs.~\cite{Yagi:2016bkt,Doneva:2017jop} for reviews on universal relations).
In the context of GW astrophysics, these imply certain binary Love relations,
presented in YY17:
\begin{enumerate}
    \item A relation between the symmetric and anti-symmetric combination of the individual
    tidal deformabilities, $\bar{\lambda}_{s} = (\lambdatilde_{1} + \lambdatilde_{2})/2$ and
    $\bar{\lambda}_{a} = (\lambdatilde_{1} - \lambdatilde_{2})/2$.
    \item A relation between the waveform tidal parameters $\bar{\Lambda}$ and $\delta \bar{\Lambda}$
    appearing in 5-PN and 6-PN order respectively.
    \item A relation between the coefficients of the Taylor expansion of the tidal deformability
    $\lambdatilde(M)$ about some mass $m_0$ as a function of mass.
\end{enumerate}
Here, we are concerned with the third item in this list, and we will refer to it as
the $\lambdazero - \lambdak$ relation.

The central idea of the $\lambdazero - \lambdak$ relation is to express $\lambdatilde(M)$ in
terms of a Taylor expansion
about a fiducial value, $M=m_0$ as, 
\begin{equation}
    \lambdatilde(M) = \sum_{k=0}^{\infty} \frac{\lambdatilde^{(k)}_{0}}{k!}\left( 1 - \frac{M}{m_0}\right)^k,
    \label{eq:lambda_m}
\end{equation}
where, $\lambdak = (-1)^{k} d^{k}\lambdatilde/{dM^{k}}$ is evaluated at $M=m_0$. The
$\lambdazero - \lambdak$ {relation} states that $\lambdak$ can be written entirely in terms of
$\lambdazero$, and the resulting expression is insensitive to the EoS. (to fractional accuracy of 
about $10\%$). 

The $\lambdazero - \lambdak$ relation can be evaluated for any tabulated EoS, but as a Fermi estimate, we
first consider the simplified example of a polytropic EoS.  For such an EoS, the pressure ($p$) and energy
density ($\epsilon$) are related via $p \propto \epsilon^{1 + 1/n}$, where $n$ is the polytropic index. To
leading order in compactness, the tidal deformability is then given by~\cite{Yagi_2013_PRD}, $\lambdatilde \propto M^{-10/(3-n)}$. The
$\lambdazero - \lambdak$ relation in this case then becomes, $\lambdak = G_{n,k}\lambdazero$, where we have defined
\begin{equation}
    G_{n,k} = \frac{\Gamma\left(k + \frac{10}{3-n}\right)}{\Gamma\left(\frac{10}{3-n}\right)}.
    \label{eq:Gnk}
\end{equation}
Thus, the binary-love-relations ensure that
$\lambdatilde^{(k)}_0 = \lambdatilde^{(k)}_0(\lambdatilde_0^{(0)})$ for $k>1$, with the relation only dependent on $n$. 

In the relativistic case and for more realistic EoSs, the tidal deformability is obtained numerically, and the
$\lambdazero - \lambdak$ relation is obtained by fitting the numerical simulations to an expression of the form (see
Eq.~(22) of YY17),
\begin{equation}
    \lambdak = G_{\bar{n},k}\;\lambdazero\left(1 + \sum_{i=1}^3a_{i,k}x^i\right),
    \label{eq:lambda_k_relativistic}
\end{equation}
where $x = (\lambdazero)^{1/5}$,  $a_{i,k}$ are numerical coefficients, and $G_{\bar{n},k}$ is
defined in Eq.~\eqref{eq:Gnk}.  Here, we will only consider an expansion to $k=3$. This
is sufficient to accurately represent $\lambdatilde(M)$  in the range $M \in (1.2\msun, 1.5\msun)$
for $m_0 = 1.4 M_\odot$, which we will choose for the rest of this work. The degree of universality
deteriorates as $k$ increases, but the overall sum is still EoS insensitive to about $10\%$ for
the mass range and $m_0$ mentioned. 

\begin{table}[]
    \centering
    \begin{tabular}{l|c|c|c}
    \hline
    \hline
         k          &   1    &    2     &    3    \\
\hline
$a_{1, k}$ & $-0.349$ &  $-1.063$  &  $-1.820$ \\
$a_{2, k}$ & 5.674  &  10.649  &  14.693 \\
$a_{3, k}$ & 0.296  &  1.800   &  2.741  \\
    \hline
    \end{tabular}
    \caption{Coefficients found by fitting the function in
    Eq.~(\ref{eq:lambda_k_relativistic}) to the $\lambdak -\lambdazero$
    {relation} data calculated from the {\numeos} EoSs we considered
    in this work. See middle (lower) panel of Fig.~\ref{fig:mass_radius_relations}
    for the fit (residuals).} \label{tab:fitting_coefficients}
\end{table}

We re-derive the fitting coefficients $a_{i,k}$ for a set of EoSs that are consistent with recent LIGO/Virgo and NICER observations (and the observation of $\sim 2M_\odot$ neutron stars~\cite{1.97NS,2.01NS,Cromartie:2019kug}). For simplicity, we here consider the piecewise polytropic representation of the EoS~\citep{Read_2009}, restricting attention only to those that have support in the $68\%$ confidence region in terms of mass and radii reported by LIGO/Virgo in Ref.~\citep{2018PhRvL.121p1101A}, and NICER in Ref.~\citep{2019ApJ...887L..24M}. 
The {upper and middle} panels of Fig.~\ref{fig:mass_radius_relations} show the mass-radius curves and the $\lambdazero - \lambdak$ relation constructed with the {\numeos} EoSs used in this work. Using Eq.~(\ref{eq:lambda_k_relativistic}) to fit for the $a_{i,k}$ coefficients up to $k=3$ yields the coefficients listed in Table~\ref{tab:fitting_coefficients}. These fitting functions give us an \emph{EoS averaged} $\lambdazero - \lambdak$ relation, as one can see from the middle panel of  Fig.~\ref{fig:mass_radius_relations}. Note that $\lambdazero$ is still a free parameter
that is to be constrained by observational data. We will here use this fit to parameterize the tidal
deformability and perform Bayesian parameter estimation on GW170817 data to measure the
constant $\lambdazero$. 

\section{$\lambdazero$ from GW170817}
\label{sec:pe}

The first step in estimating the Hubble parameter with the binary Love relations is to estimate
$\lambdazero$ with a controlled event. We provide a brief review of Bayesian inference.
The aim of Bayesian inference is to obtain a posterior probability density for the
parameters describing the signal, $\pmb{\Theta}$, given the GW data, $\xgw$,
\begin{equation}
    p(\pmb{\Theta}\vert\xgw) \propto p(\xgw \vert\pmb{\Theta})\; p(\pmb{\Theta}).
\end{equation}
Here, $p(\xgw \vert\pmb{\Theta})$ is the likelihood of the parameters and $p(\pmb{\Theta})$
is the prior distribution. Gravitational waves from compact binary coalescences of BBHs are
described by 15 parameters. These contain the intrinsic parameters like
the masses ($m_{1,2}$) and spin parameter 3-vectors ($\pmb{a}_{1,2}$), and the extrinsic parameters like luminosity
distance, $\dl$, inclination angle, $\iota$, and so on (see, for example,
Refs.~\cite{PhysRevD.49.1723,Cutler_1994,van_der_Sluys_2008,Veitch_2012,Veitch_2015}).
In the case of BNSs, matter effects enter the waveform phase first at 5PN and then 6PN
order via two additional tidal deformability parameters, $\lambdatilde_{1,2}$, as,
\begin{eqnarray}
    \Psi_{\text{tid}} &=& -\frac{3}{128\eta x^{5/2}}
        \left[
        \frac{39}{2}\bar{\Lambda}x^5 + \right. \nonumber \\
        &&\left. \left(\frac{3115}{64}\bar{\Lambda} - 
        \frac{6595}{364}\sqrt{1 - 4\eta}\delta\bar{\Lambda}\right)x^6 + \mathcal{O}(x^7)
        \right],
\end{eqnarray}
where $x = \left[\pi(m_1 + m_2)f\right]^{2/3}$ is the PN expansion parameter, $f$ is the
GW frequency, and $\eta = m_1m_2/(m_1 + m_2)^2$ is the symmetric mass-ratio. The expressions
for $\bar{\Lambda}$ and $\delta\bar{\Lambda}$ have the form,
\begin{eqnarray}
    \bar{\Lambda} &=& f(\eta)\left(\frac{\lambdatilde_1 + \lambdatilde_2}{2}\right) 
        + g(\eta)\left(\frac{\lambdatilde_1 - \lambdatilde_2}{2}\right) \nonumber \\
    \delta\bar{\Lambda} &=& \delta f(\eta)\left(\frac{\lambdatilde_1 + \lambdatilde_2}{2}\right) 
        + \delta g(\eta)\left(\frac{\lambdatilde_1 - \lambdatilde_2}{2}\right),
\end{eqnarray}
where the exact expressions for $\{f, g, \delta f, \delta g\}$ are given in Sec.~2.2 of YY17. In our case,
we are interested in measuring the tidal deformability by using the
parameterization described in Eq.~(\ref{eq:lambda_m}) and Eq.~(\ref{eq:lambda_k_relativistic}), but of course, all waveform parameters must be searched over when exploring the likelihood surface.

\begin{figure}[t]
    \centering
    \includegraphics[width=1.0\columnwidth, trim=0cm 1cm 0cm 0cm]{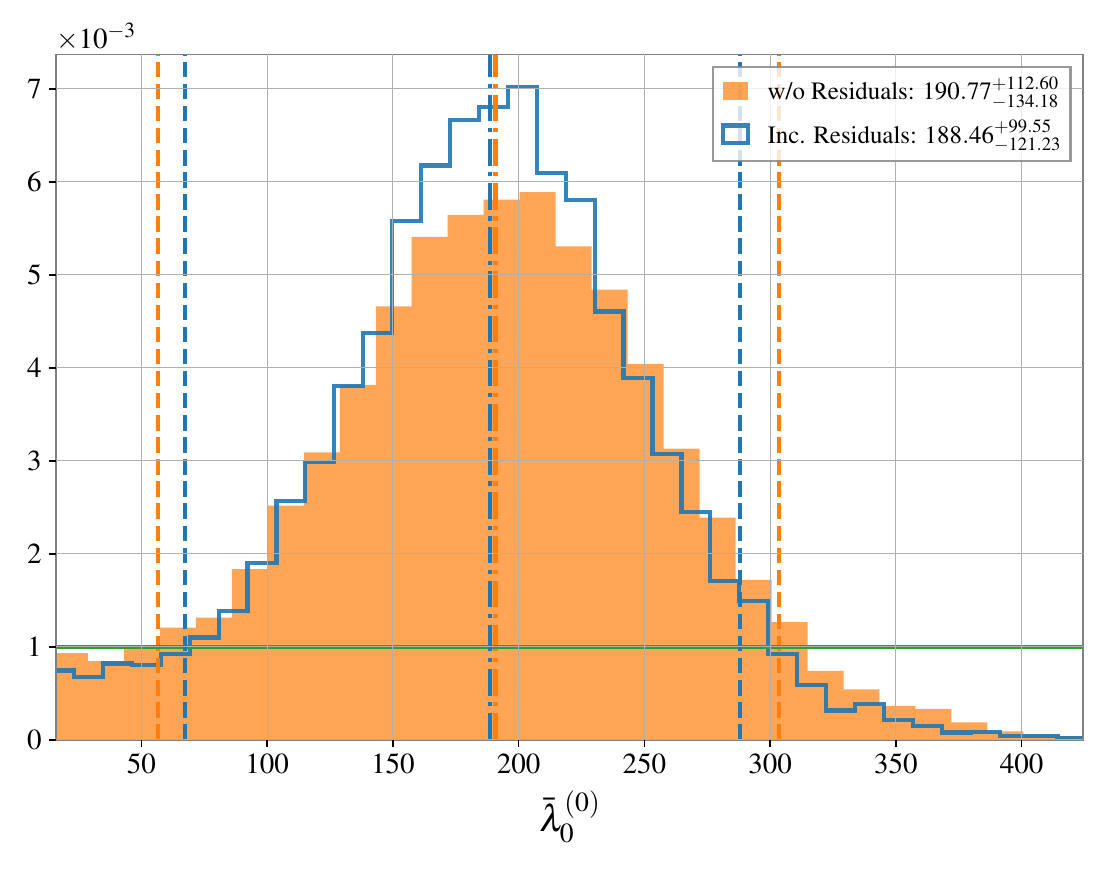}
    \caption{
    Posterior probability density for the $\lambdazero$ parameter measured from 128s of
    strain data (4~kHz) for GW170817. The $\lambdazero-\lambdak$ relations are used in the parameterization of the tidal
    deformability in Eq.~(\ref{eq:lambda_m}) with (unfilled) and without (filled) the residual errors
    in the relations. The redshift is fixed at the host galaxy $\zngc=0.0099$, and
    therefore, the tidal deformability is parameterized only by $\lambdazero$. The vertical 
    dash-dot lines represent the median, the vertical dashed lines represent the 90\% confidence
    interval, and the horizontal line is the prior density.
    }
    \label{fig:lambda_zero_posterior}
\end{figure}
To obtain a measurement for $\lambdazero$, we perform Bayesian parameter estimation
on 128s of 4~kHz strain data for GW170817.\footnote{Taken from Gravitational-wave Open
Science Center (GWOSC) \url{https://www.gw-openscience.org/catalog/GWTC-1-confident/single/GW170817/}}
For the model, we use the IMRPhenomPv2\_NRTidal waveform~\cite{PhysRevD.99.024029}, which
is the same as that used for parameter estimation in Ref.~\cite{2019PhRvX...9a1001A} and
Ref.~\cite{Abbott_2019} (hereafter LV19).\footnote{Both references analyze the same duration
of data with the same settings, but the latter uses data with a different calibration. The
data from the latter is available on GWOSC, which is what we use here.} 
In our analysis, however,  the parameters $\bar \lambda_{1,2}$ are not independent, but rather
they are modeled through the Taylor expansion in
Eq.~\eqref{eq:lambda_m}, with the EoS-insensitive $\lambdazero-\lambdak$ relation of
Eq.~\eqref{eq:lambda_k_relativistic}. This then means that
$(\bar{\Lambda},\delta\bar{\Lambda})$ (or equivalently $(\lambdatilde_{1},\lambdatilde_{2})$) are
not parameters of our model anymore, but rather the model now only depends on a \textit{single} tidal
deformability parameter, namely $\lambdazero$. The model also depends on the detector-frame masses
${\mdet}{}_{1,2}$ and the redshift $z$, since the Taylor expansion of Eq.~\eqref{eq:lambda_m}
depends on the source-frame masses ${\msource}{}_{1,2}$. For the GW170817 event, however, the
redshift is known to high accuracy due to identification of the host galaxy, and so this is no
longer a free parameter in this case. 

The priors on the parameters of our model are chosen as follows. For the extrinsic parameters, like the
distance, inclination, coalescence phase, sky-position and so on we pick uninformative priors as
those in LV19.\footnote{In LV19, the sky-position was fixed to the location of the host galaxy.
Here, we leave them free too.}
Following LV19, we sample on detector-frame component masses setting the prior distribution of each component mass as
the marginalized posterior distribution obtained for the component masses from LV19. We also
repeat the analysis putting a flat prior on the component masses as in LV19, and find similar results.
For the spins, we use the low-spin prior from LV19 i.e., $\chi_i \leq 0.05$. For  $\lambdazero$, we use
a flat prior in $[0, 1000]$, which implies a prior on the individual tidal deformabilities through
Eq.~(\ref{eq:lambda_m}) for a fixed redshift. The implied priors on the individual tidal deformabilities
are shown in Appendix~\ref{appendix:lambda_1_2_prior}. This step is different from LV19, where a uniform prior was used
on the individual tidal deformabilities, $\lambdatilde_{1,2}$, independently. We fix the redshift of the source at the
value of the host galaxy NGC4993, $\zngc = 0.0099$, which was determined through EM observations reported
in Ref.~\cite{Levan_2017}, and used in LV19. This allows us to obtain the source-frame component masses,
${\msource}_{1,2} = {\mdet}_{1,2} /(1 + \zngc)$. 
The setup of the likelihood function and the sampling is done through the open-source GW parameter
estimation library, {\bilby},~\citep{bilby} using with the adaptive nested sampler, {\dynesty}~\citep{dynesty}.
The resulting marginalized distribution for $\lambdazero$ is shown in Fig.~\ref{fig:lambda_zero_posterior}.
The distribution of the other parameters is presented in Appendix~\ref{appendix:pe}.

We find the value of
$\lambdazero = 191^{+113}_{-134}$ at 90\% confidence. In order to verify the measurement is robust
against the residual errors coming from the $\lambdazero-\lambdak$ {relation} fit, we repeat the analysis sampling
in $\lambdazero$, but also sampling from the residual error in the $\lambdak(\lambdazero)$ relation
error, shown in the bottom panel of Fig.~\ref{fig:mass_radius_relations}. In this case, when sampling,
for each throw of a $\lambdazero$ sample, we obtain the $\lambdak$ using the fit and a sample from the
distribution of residuals.
The posterior density of $\lambdazero$ obtained this way is shown
in Fig.~\ref{fig:lambda_zero_posterior} using the un-filled histogram. We find that the result is not significantly
affected ($\sim 2\%$ difference in median values).\footnote{Kullback–Leibler divergence is $\mathcal{O}(10^{-2})$,
regardless of the direction of the comparison.}
Our results are consistent with the measurement of $\lambdatilde(1.4\msun) = 190^{+390}_{-120}$
obtained by a linear expansion of $\lambdatilde(\msource)\msource^5$ about $1.4~\msun$, reported in
Ref.~\cite{2018PhRvL.121p1101A} following the approach in Refs.~\citep{delpozzo_2013,agathos_2015}.
The difference in the upper limit is caused due to differences in prior implied on the tidal
deformabilities. In Appendix~\ref{appendix:lambda_1_2_prior}, we compare our prior with that in
Ref.~\cite{2018PhRvL.121p1101A}.

The first measurement of $\lambdazero$ reported above is perhaps not extremely constraining, but one
expects future events to allow for more accurate measurements. Future observing runs of LIGO/Virgo/KAGRA
with coincident operation of next generation telescope facilities, like the Rubin
Observatory~\cite{Ivezic_2019}, will yield many more multi-messenger BNS
events. We can think of $\lambdazero$ as a variable which parameterises $\lambdatilde$ as
a function of the source mass as in Eq.~(\ref{eq:lambda_m}). Using the parameterization in
Eq.~(\ref{eq:lambda_m}) we can stack data from multiple observations to yield an improved
measurement of {\lambdazero}. This implies that eventually, we will be able to
``fix'' the value of $\lambdazero$, or marginalize over the small measurement (and
eventually systematic) uncertainties.

\section{Measuring {\hubble} with Love}\label{sec:hubble}
If $\lambdazero$ has been estimated by a set of controlled observations (like the GW170817 event), what else can be done with additional future observations of BNS events? Using the same parameterization in Eq.~(\ref{eq:lambda_m}), we see then that the tidal deformability terms of the waveform phase are now only a function of {\msource},
or equivalently of the detector-frame masses and the redshift. Since the detector-frame masses can be separately and tightly estimated from lower-PN order terms, the tidal deformability terms now yield information exclusively on the redshift~\cite{messenger_read_2012}. Therefore, any BNS signal, \textit{irrespective of being well-localized or
having an associated counterpart}, would result in a direct measurement of the redshift from the tidal deformability, and  this can be used to infer the Hubble 
constant, {\hubble}.

Let us then consider the prospects of measuring $\hubble$ with future observations, beginning with a best-case
scenario, in which we assume $\lambdazero$ has been strongly constrained. We re-write Eq.~(\ref{eq:lambda_m}) as,
\begin{equation}
    \lambdatilde_{1,2} = \lambdazero + \sum_{k=1}^N 
        \frac{\lambdak}{k!} \left(1 - \frac{\mdet{}_{1,2}/m_0}{1 + z(\dl, \hubble, \bomega)}\right)^k,
    \label{eq:lambda_h0}
\end{equation}
where $\bomega = \{\Omega_i\}$ are the cosmological parameters apart from \hubble, which we here, for simplicity, fix to a flat {\lcdm} model with an assumed ``true'' value of {\hubble = 70\;\hubbleunit} and matter-to-critical-density ratio $\Omega_{m0}=0.3$. Hence, we can write
$z = z(\dl, \hubble)$ in Eq.~(\ref{eq:lambda_h0}). The measurement of {\dl}
comes from the waveform amplitude, and thus the redshift, $z$, alone comes from Eq.~(\ref{eq:lambda_h0}). This enables either direct
sampling of {\hubble}, or we can infer {\hubble} in a post-processing step after we have sampled in $z$.

Even though the logic behind this idea is straightforward and robust, its implementation for a single event is 
hindered by measurement uncertainties and covariances. In particular, the distance-inclination degeneracy
in the amplitude implies that distance measurements peak at lower-than-true
values for face-on sources, and to greater-than-true values for edge-on sources
(see Refs.~\cite{2020PhRvD.101l2001G,Chen_2018}, for example).
Thus, {\hubble} measurements from individual events may be multi-modal and, in general,
peaked away from the true value. However, since all observations should depend on the same $H_0$ (assuming this quantity is truly a constant),
one should be able to \emph{stack} multiple events to obtain an accurate measurement of the Hubble constant. 
\begin{figure*}[!thp]
    \centering
    \includegraphics[width=0.48\textwidth, trim=0cm 0cm 0cm 0cm]{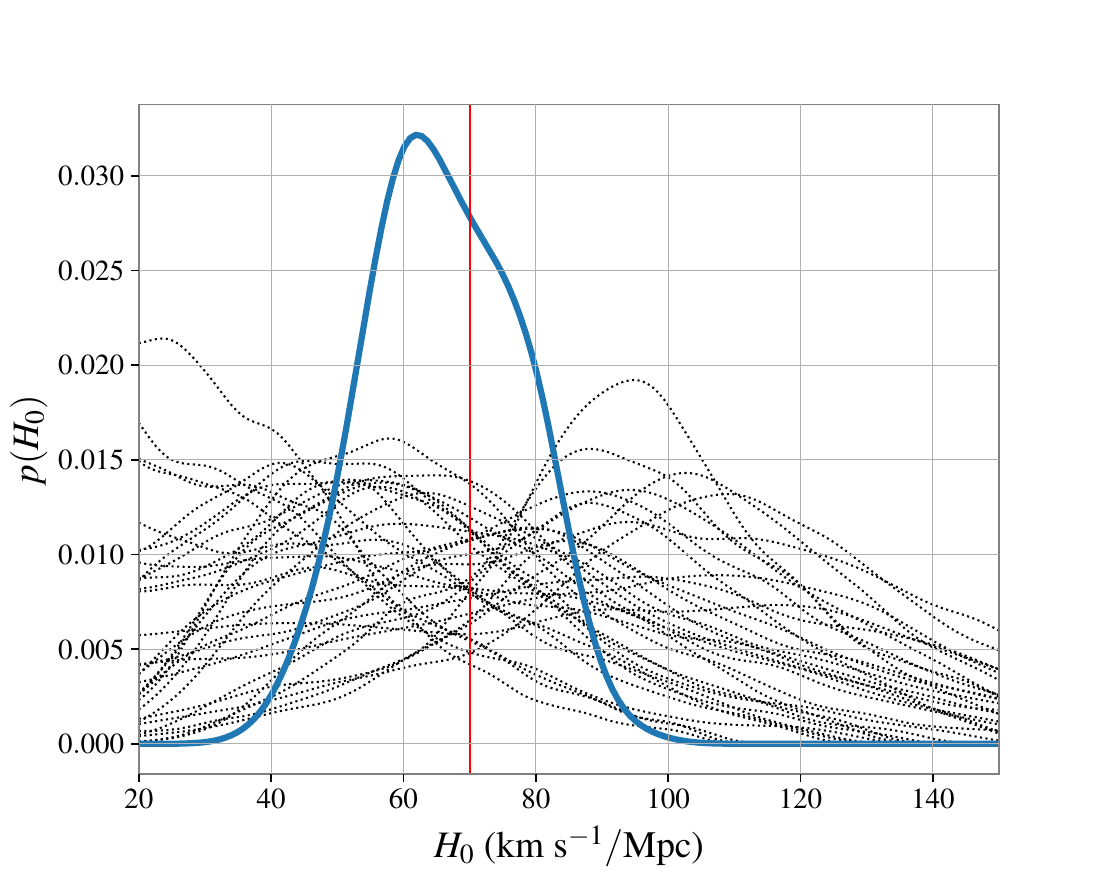}
    \includegraphics[width=0.48\textwidth, trim=0cm 0cm 0cm 0cm]{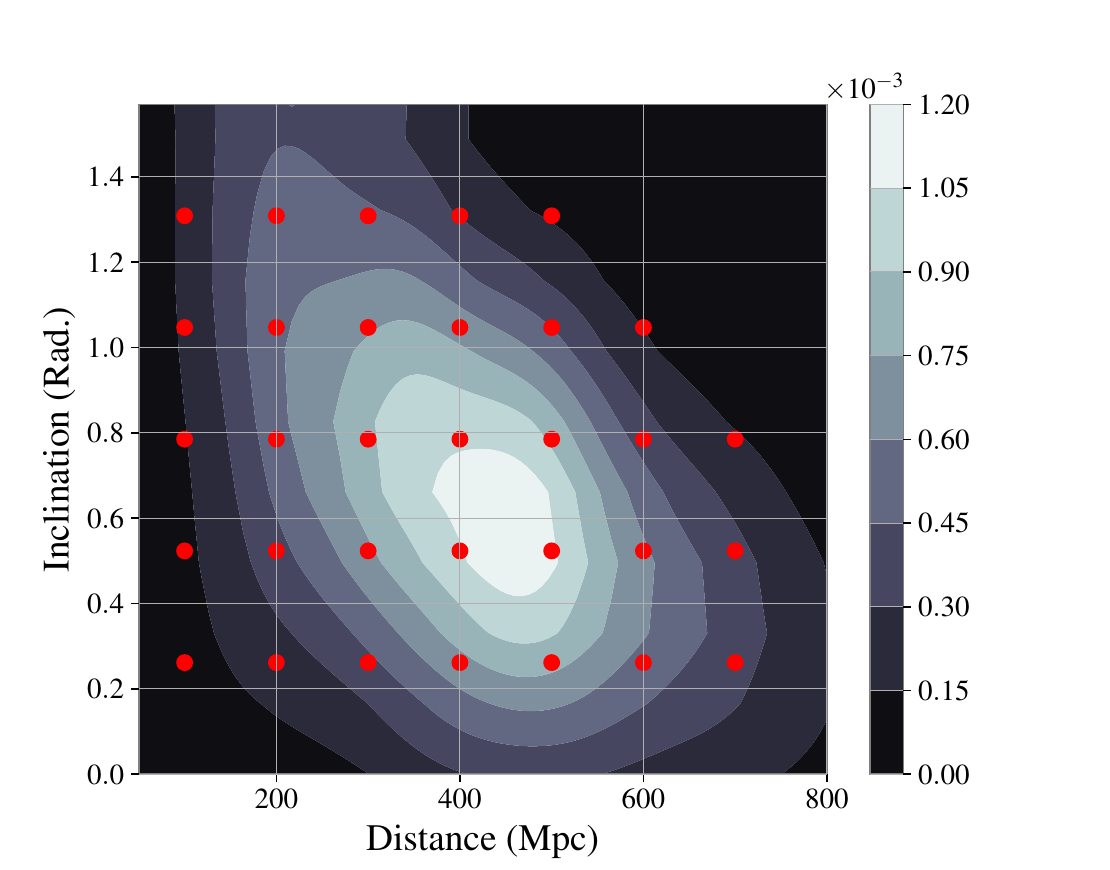}\\
    \includegraphics[width=0.48\textwidth, trim=0cm 0cm 0cm 0cm]{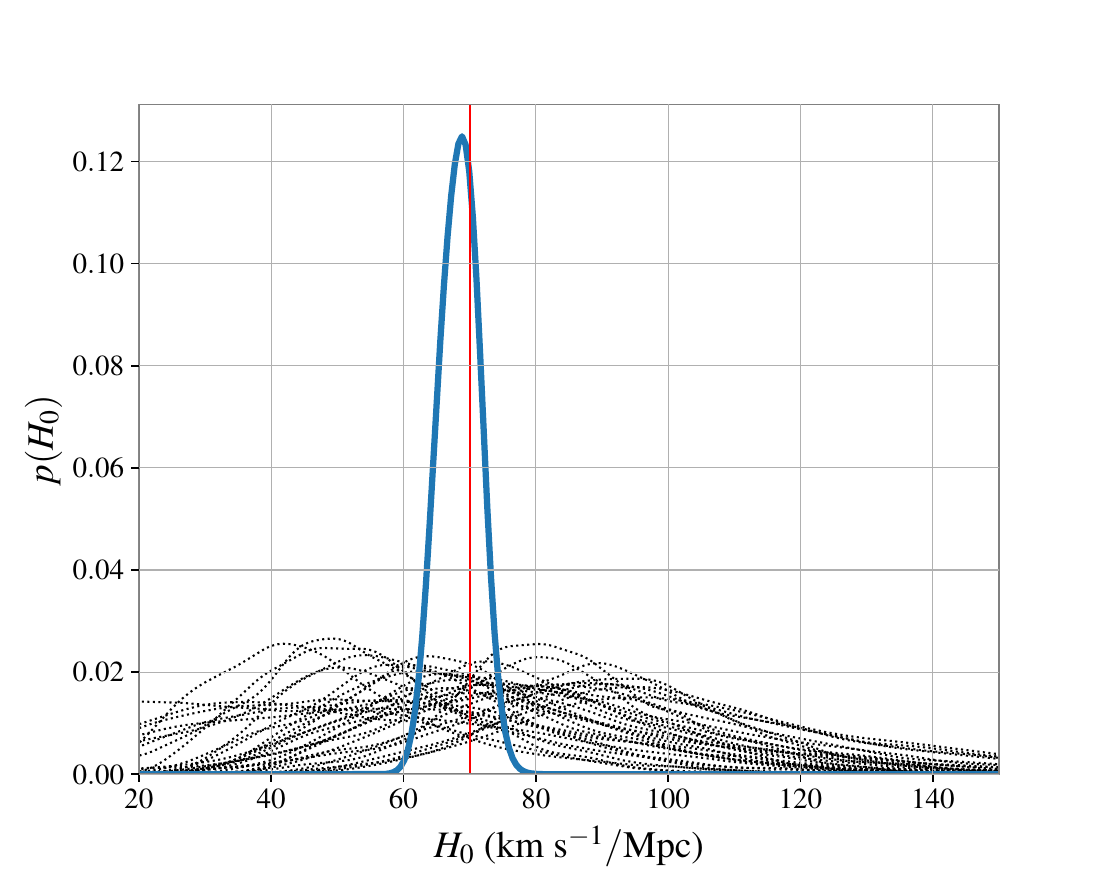}
    \includegraphics[width=0.48\textwidth, trim=0cm 0cm 0cm 0cm]{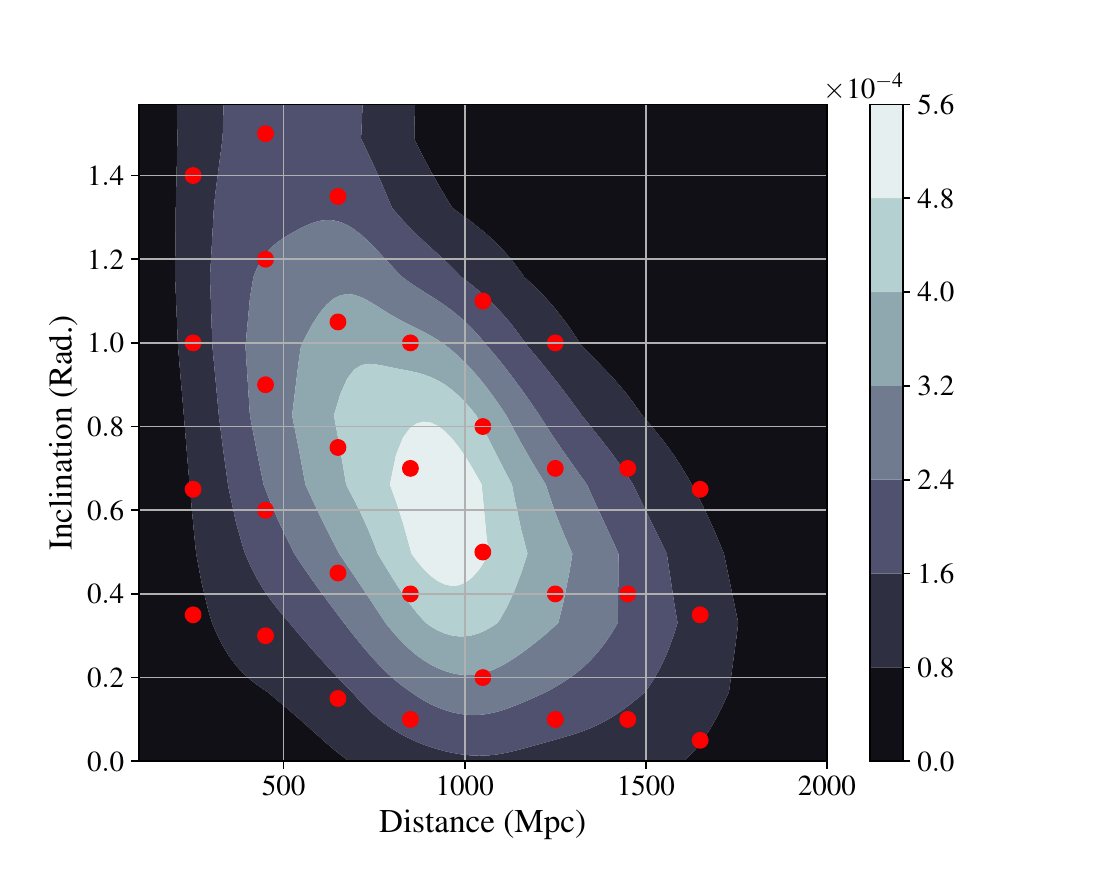}\\
    \includegraphics[width=0.48\textwidth, trim=0cm 0cm 0cm 0cm]{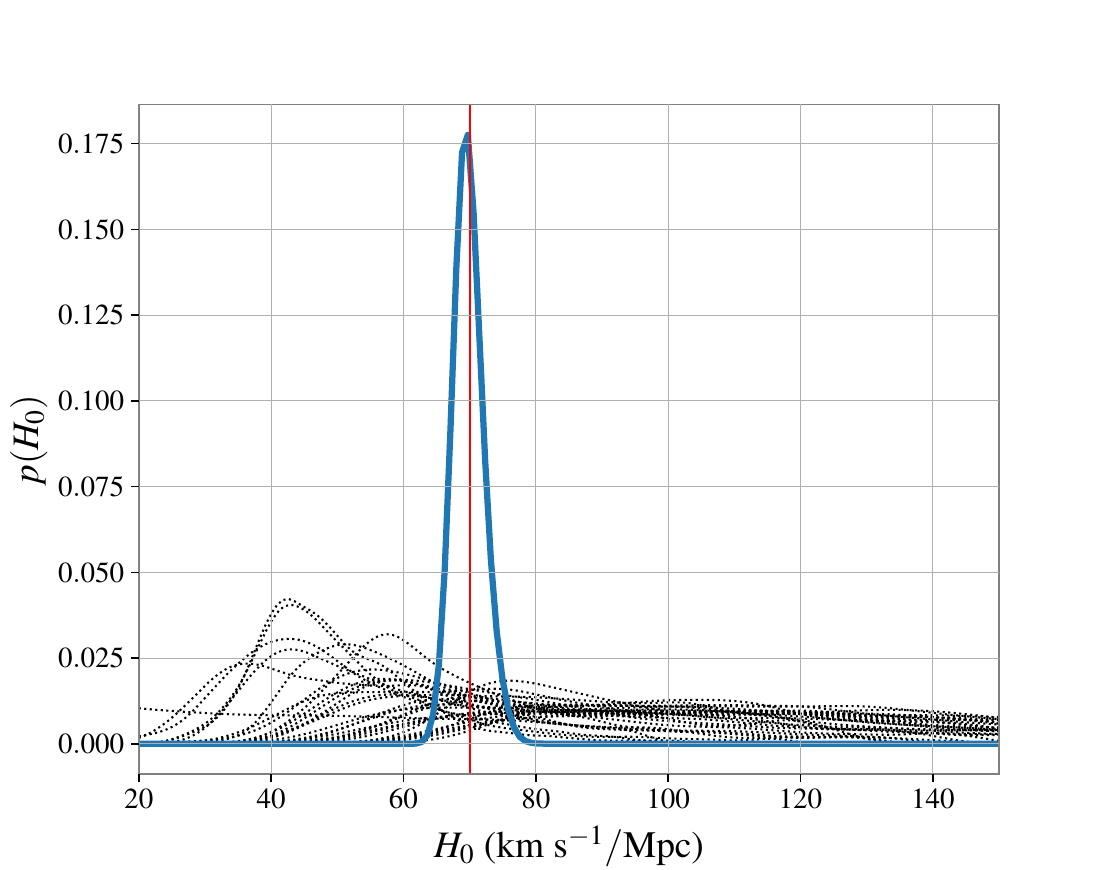}
    \includegraphics[width=0.48\textwidth, trim=0cm 0cm 0cm 0cm]{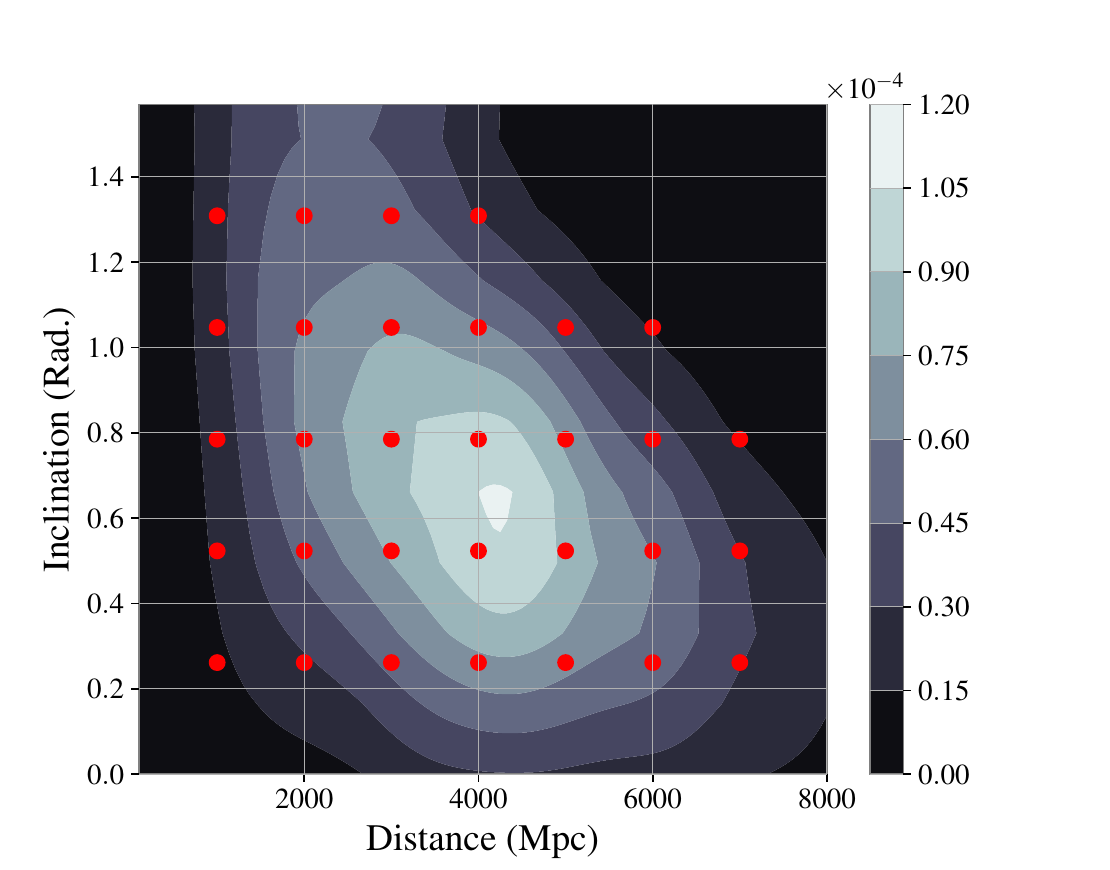} \\

    \caption{{\bf{Left}}: Individual (dotted) and stacked (solid) {\hubble} likelihoods
    obtained for the representative runs in the O5 (top), Voyager (middle) and CE (bottom)
    observing eras. 
    Due to distance-inclination degeneracy, the individual posteriors are sometimes peaked
    away from the true value. However, after stacking the posteriors based on a sample of
    {\Ndet} events from the detected population, the distribution
    converges to the true simulated value, shown by the vertical line, $\hubble = 70 \hubbleunit$. 
    For illustration, we choose {\Ndet=10, 50, 70} for the top middle, and bottom panels,
    corresponding to the O5, Voyager, and CE era, respectively.
    {\textbf{Right:} The normalized number count of of distances and inclinations of the systems
    that would be detectable in a specific observing era is shown by the contour map. This
    was obtained by simulating an ensemble (several thousands) of binaries with
    $(m_1, m_2) \sim (1.42, 1.27)~M_{\odot}$ distribution uniform in sky, volume, and inclination,
    and requiring that each produce network SNR $\geq 8$. The grid of
    overlayed points represents the true distance/inclination of synthetic systems
    chosen for performing Bayesian parameter estimation. We observe that the {\hubble}
    measurement converges irrespective of the shape of grid as long as it covers
    the distances and inclinations that would be detected.
    } 
    }
    \label{fig:stacked_h0}
\end{figure*}

We investigate how this stacking measurement could take place in the future. We
simulate synthetic GW signals in three different networks of detectors:
ground-based observatories in the O5 era (HLVKI 5 detector network), the Voyager era
(HLI upgraded to Voyager + VK), and the Cosmic Explorer era (1 CE instrument at H).
For simplicity, we consider a set of BNS signals produced by systems with a fixed source frame
chirp mass, $\mchirp = 1.17~\msun$, and mass-ratio, $q=0.9$, but at different distances, sky locations
and inclination angles. We assume the true EoS is such that it would lead to a tidal deformability
of $\bar{\lambda}(1.4~\msun) = \lambdazero = 200$ for a $1.4 M_\odot$ NS, a value motivated from the
previous section. 

The detectability of our synthetic catalog of signals is estimated as follows. We first create an
ensemble (several thousands) of such fiducial systems, distributed uniformly in sky-location, orientation,
and volume {($\propto \dl^2$)}.
We then say that a fiducial system can be detected if the network signal-to-noise ratio (SNR)
is above 8 using the appropriate noise spectral densities for the detectors of in each era. For the O5 era,
we use the noise curve from Ref.~\cite{observing_scenarios}.~\footnote{\url{https://dcc.ligo.org/LIGO-T2000012/public}}
For the Voyager and CE era, we use noise curves from Ref.~\cite{Abbott_2017}.~\footnote{\url{https://dcc.ligo.org/LIGO-T1500293/public}}
Some of these noise curves are also present as package data in {\bilby}.
The distribution of detected distances and inclination angles are shown in the right panels of
Fig.~\ref{fig:stacked_h0} for each era. We can see that the detected distance distribution is peaked
at a certain distance depending on the era, due to the combination of the prior distribution and the detector
sensitivity. We also see the preference towards detecting face-on sources, as opposed to edge-on sources at the
same distance. However, the marginalized distribution of detected inclination angles are universal (see Fig.~4
of Ref.~\cite{schutz_2011}). We have verified this for each of our eras.
We focus on distance and inclination because of the distance-inclination degeneracy during
parameter recovery. The detectability, in general, also depends on the mass-distribution of BNSs in the universe,
which we ignore for this work. But, this dependence is weak due to the narrow distribution of BNS
masses~\citep{Chen_2018,Fishbach_2019}.
In Appendix~\ref{appendix:hubble-trends}, we show effect on distance recovery based on the
true inclination angle of the source.

During an observing run, we expect a sample of detected BNS events, {\Ndet}, whose true values of
distance and inclination are based on this distribution. In reality, the value of {\Ndet} will depend
on the volumetric rate of BNS mergers, and also the redshift evolution of the rate.
Here, we will simplify the problem, considering the following representative cases 
$\Ndet = 10^1, 10^2$, and $10^3$ corresponding to expected numbers for the O5, Voyager,
and CE detector eras i.e., {\Ndet} events are drawn from the distributions shown in
Fig.~\ref{fig:stacked_h0}. Since the computational cost of performing Bayesian parameter
estimation on all {\Ndet} systems is high, we instead simulate the fiducial source at certain
representative points on a grid of distances and inclination angles. These are represented by
the solid points overlayed on the heatmap in Fig.~\ref{fig:stacked_h0}. We then count each
representative run based on the relative probability of the heatmap, $p_{\text{det}}$, normalizing
the total count to {\Ndet}.
Thus, {\Ndet} is approximated as,
\begin{equation}
    \Ndet = \sum_{\alpha} N_{\alpha},
\end{equation}
where the summation index runs over all the grid points, and
$N_{\alpha} \propto p_{\text{det}}(D_L^{\alpha}, \iota^{\alpha})$.
{In Appendix~\ref{appendix:alternative_prior}, we consider
alternate physically motivated distance distributions from different
rate models. We find that the choice does not affect the answer.}

The combined {\hubble} posterior is given by,
\begin{eqnarray}
    p(\hubble\vert \{d_1,\dots,d_{\Ndet}\}) \propto \hspace{4cm}\nonumber \\
    p(\hubble)\prod_{i=1}^{\Ndet}~\frac{1}{\beta(\hubble)}\times
    \int \bigg[ d\pmb{\Theta}_i\;\delta[z - \hat{z}(\dl, \hubble)]\hspace{1cm}
    \nonumber \\
    p(d_i\vert \underbrace{
        \mdet{}_{1,2}, \pmb{a}_{1,2}, \dots, \dl, z, \lambdazero
    }_{\pmb{\Theta}_i})~p(\pmb{\Theta}_i)\bigg]\label{eq:h0_likelihood1}~\\
    \propto p(\hubble)\prod_{i=1}^{\Ndet}~{\cal{L}}_i(\hubble)\hspace{3cm}
    \label{eq:h0_likelihood1.1}
\end{eqnarray}
where $\pmb{\Theta}_i = \{\mdet{}_{1,2}, \pmb{a}_{1,2}, \dots, \dl, z, \lambdazero\}_i$
is the set of GW parameters for an individual event now parameterized by $\lambdazero$
and $z$ in place of $\lambdatilde_{1,2}$. The likelihood function of a single event
is given by $p(d_i\vert \pmb{\Theta}_i)$. The prior distributions for individual
parameters are given by $p(\pmb{\Theta}_i)$.
The constraint between $\dl$ and $z$ is represented by the $\delta$-function term. With this
constraint, marginalizing over all parameters $\pmb{\Theta}_i$, leads to a posterior of
{\hubble} from an individual event. Although, a explicit prior on {\hubble} is not applied,
any reasonable prior on the distance and redshift for the individual events results in an
implied prior on {\hubble},
\begin{equation}
    \beta(\hubble) \propto \int d\pmb{\Theta}_i\;p(\pmb{\Theta}_i)\;\delta[z - \hat{z}(\dl, \hubble)].
    \label{eq:beta_definition}
\end{equation}
By ``dividing-out'' this prior in Eq.~(\ref{eq:h0_likelihood1}), we get
the individual semi-marginalized likelihoods ${\cal{L}}_i(\hubble)$. These are
then  multiplied together to obtain the joint {\hubble} likelihood, and the 
stacked posterior in Eq.~(\ref{eq:h0_likelihood1.1}).
This ``dividing-out'' procedure guarantees that in the absence of any signal,
the likelihood is flat (see Appendix ~\ref{appendix:reweighting}).
Equation~\eqref{eq:h0_likelihood1.1} can be simplified through our counting method to obtain
\begin{align}
    &p(\hubble\vert \{d_1,\dots,d_{\Ndet}\}) \propto 
    p(\hubble) 
    \nonumber \\
    &\qquad  \times\left[ 
    \left(\underbrace{\mathcal{L}_{\alpha=1}(\hubble)\times\dots}_{N_{\alpha=1}~\text{times}}\right)
    \times
    \left(\underbrace{\mathcal{L}_{\alpha=2}(\hubble)\times\dots}_{N_{\alpha=2}~\text{times}}\right)
    \times\dots
    \right], \nonumber \\
    \label{eq:h0_likelihood2}
\end{align}
where $\alpha=1, 2,\dots$ goes over the
representative grid points in distance and inclination, and each likelihood is counted based on its relative probability of detection. In practice, the implementation of Eq.~\eqref{eq:h0_likelihood2} is simpler, so we adopt it henceforth. 

With all of this at hand, let us now estimate the accuracy to which the Hubble parameter
could be inferred in the future as follows. For each of the representative grid points,
we inject a non-spinning waveform corresponding to the fiducial system mentioned above
in a noise realization based on the observing scenario. We then perform Bayesian parameter
estimation using the {\pbilby} inference library~\cite{Smith_2020}, and the same
waveform model, IMRPhenomPv2\_NRTidal, for both injection and recovery.
We inject BNS waveforms represented at the grid points above in a noise-realization.
We sample in the detector-frame chirp mass and mass-ratio, while ignoring spins since they have a negligible
effect in our analysis. We also keep the sky-location fixed to the injected value for simplicity. We have
checked that setting these free does not impact the result. We also fix the coalescence time to the injected
value. This is motivated since in practice GW compact binary search pipelines report the coalescence
time. We put a uniform in comoving-volume prior for the luminosity distance, and a uniform in cosine prior for
the inclination angle. We also sample in the redshift, $z$, with a uniform prior, convert the detector-frame
quantities to source-frame, and use the $\lambdazero-\lambdak$ relations to obtain the tidal deformabilities.
More details about {\pbilby} settings are give in Appendix~\ref{appendix:pbilby}.

Given the analysis described above at each grid point, we then obtain the individual {\hubble} posteriors as a
post-processing step. First, we divide out the individual likelihoods by any implied prior due to the $\dl$
and $z$ prior combination mentioned in Eq.~(\ref{eq:beta_definition}) to obtain the likelihoods $\cal{L}_{\alpha}$
in Eq.~(\ref{eq:h0_likelihood2}). We have repeated this analysis by sampling directly on {\hubble} and using a
flat prior on this quantity, in which case the additional ``dividing-out''  step is not required (in both cases,
we obtain the same results). With the individual likelihoods at hand, we then multiply the individual likelihoods
together based on their relative detection counts, normalizing to {\Ndet} events. We then use an overall flat
prior on the stacked Hubble parameter to get the stacked {\hubble} posterior.

The results are
shown in Fig.~\ref{fig:stacked_h0}, where the upper middle and lower panels represent the O5, Voyager,
and CE eras. The grid points in the right panel are the true $\dl$ and $\iota$ for which we obtain
representative PE runs, and count them based on the relative values of the heatmap at the grid
points.\footnote{Repeating the same analysis to obtain the relative counts by integrating over a small
patch around each grid point does not change our conclusions.} We use a generic choice for the
representative points for the grid, and find that our final stacked posteriors are agnostic of the choice made (observe the difference between the middle panel and the other two), since we count the relative occurrence based on
the detectability of that particular injection. The individual {\hubble} likelihoods are shown in
dashed curves in the left panels. The likelihood, after combining {\Ndet} events, is shown with a thick,
solid curve. Although the individual likelihoods may be peaked away from the true value (shown by the vertical
line), they still have support at the true value. Combining the results via stacking leads to a stacked posterior that peaks at the injected value. For illustration, in Fig.~\ref{fig:stacked_h0}, the middle and bottom panels use {$\Ndet=50, 70$} for Voyager and CE era, respectively. 
To obtain the uncertainty in the measurement of the {\hubble} constant with
{$\Ndet=10^2, 10^3$} events, we use a $\sim 1/\sqrt{\Ndet}$ scaling and find that
$\Delta\hubble/\hubble \sim 10\%$ for Voyager and $\sim~2\%$ for CE respectively.

\section{Robustness of Forecasts}{\label{sec:uncertainties}}
In the previous section, we made several assumptions to arrive at a forecast of how well {\hubble} could be measured in the future through the stacking of multiple events. In this section, we investigate the robustness of these forecasts by relaxing some of our assumptions. One of this assumptions was that $\lambdazero$ had been measured perfectly, i.e., we used a fixed delta-function prior for $\lambdazero$ peaked at the injected value. Even though this is a justified assumption as more BNS mergers with counterparts are discovered (since each individual event will constrain the value of $\lambdazero$ more and more tightly), the posterior on $\lambdazero$ will never be a delta function centered at the true value and this will deteriorate our measurements of \hubble. Another assumption was that the binary Love relations were exactly EoS independent. Although these relations are indeed EoS insensitive, they are not exactly universal, and this could lead to a systematic bias in the estimates of {\hubble}. We will investigate each of these issues in turn.  

\subsection{Effect of statistical uncertainty in {\lambdazero}}
\begin{figure}[h!]
    \centering
    \includegraphics[width=1.0\columnwidth, trim=0cm 0cm 1cm 0cm, clip]{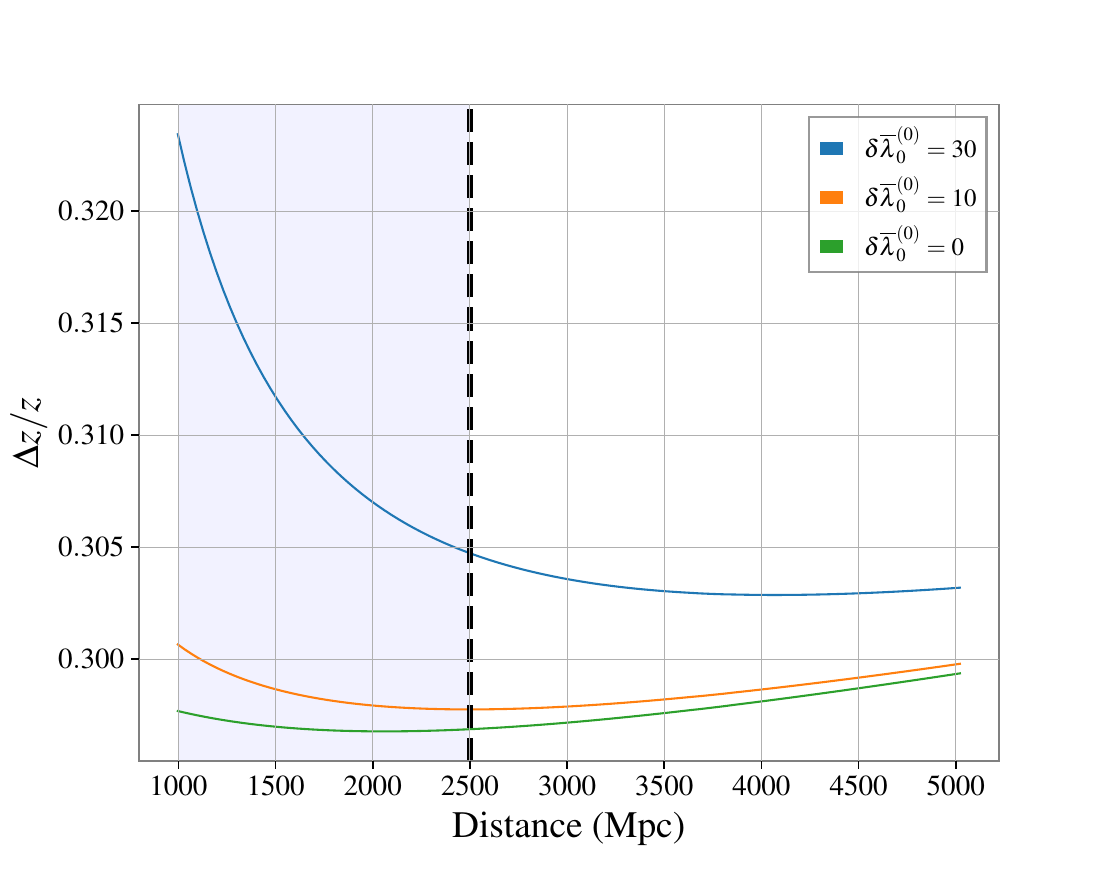}
    \includegraphics[width=1.0\columnwidth]{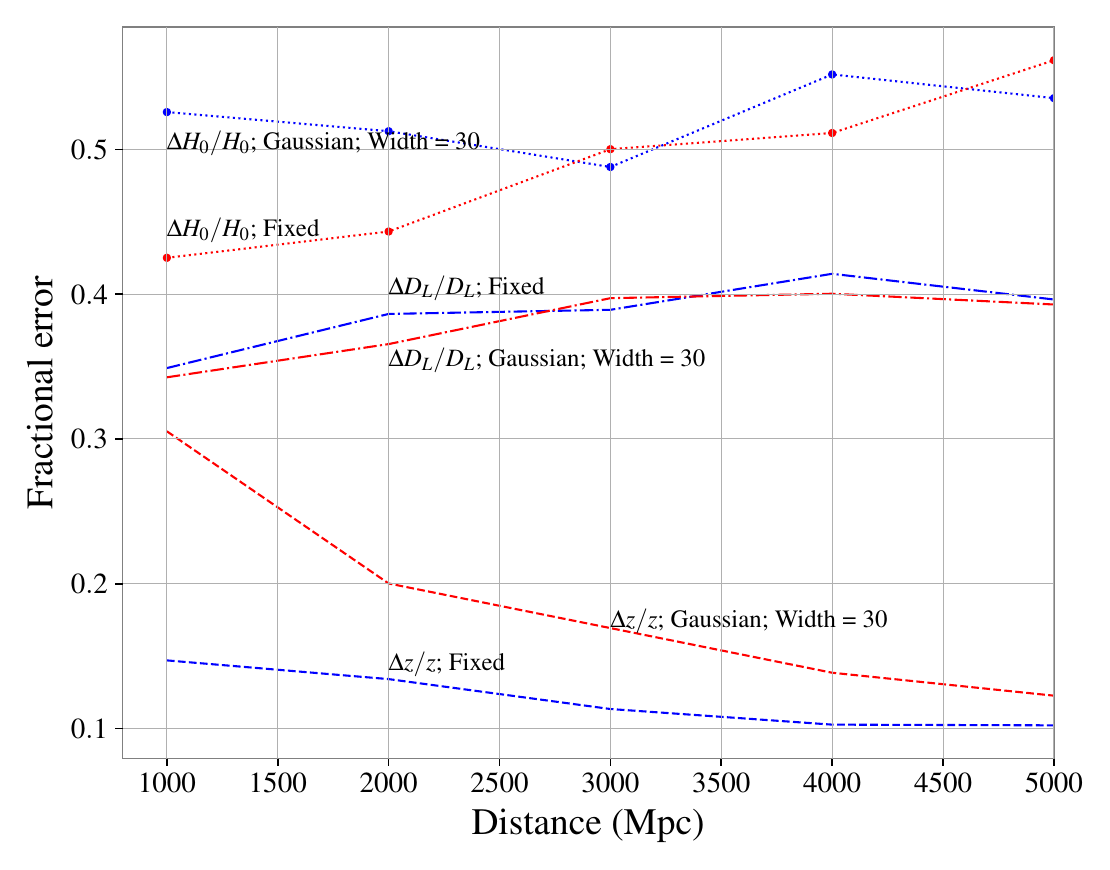}
    \caption{\textbf{Top Panel}: The fractional error in $z$ as a function of the width in $\lambdazero$ prior
    obtained using Fisher analysis on the model in Eq.~(\ref{eq:fisher_model}). We see that the fractional uncertainty in $z$
    follows the trends found in Ref.~\cite{messenger_read_2012} (analogous to the $\delta\lambdazero = 0$ case).
    We see that, the fractional error in redshift does not change significantly with change in the width of $\lambdazero$
    prior at large distances from the source.
    {The Fisher approximation is however valid in the high SNR regime, which we assume to be the region
    of SNR $\gtrsim 30$ shown by the shading.} We find similar trends from Bayesian parameter estimation results
    shown in the bottom panel.
    \textbf{Bottom Panel}: Fractional statistical uncertainty in the measurement of the redshift, $z$, the
    luminosity distance, $\dl$, and {\hubble} when using a Gaussian prior on $\lambdazero$ versus when using
    a $\delta$-function prior (fixed) are shown for a few representative parameter-estimation runs in the CE
    era.
    }
    \label{fig:h0_uncertainty_gaussian_prior}
\end{figure}
Equation~(\ref{eq:lambda_h0}) tells us that the effect of statistical measurement uncertainty in
{\lambdazero} directly affects the measurement of the redshift, $z$. To estimate this effect, we perform a
Fisher analysis similar to that of Ref.~\cite{messenger_read_2012}. 
For the signal model we use a restricted post-Newtonian(PN) waveform, where we include terms up to 3.5 PN for
the point particle contribution \cite{PhysRevD.71.084008} and upto 7.5 PN in the tidal contribution in the phase
of the waveform \cite{damour_2012}. 
We parametrise our waveform as 
\begin{equation}
    \tilde{h}(f) = \tilde{h}(f; \mathcal{A}, \mathcal{M}_c, \mathcal{\eta}, \dl, t_c, \phi_c, \lambdazero, z),
    \label{eq:fisher_model}
\end{equation}
where $t_c$ and $\phi_c$ are the coalescence time and phase and $\mathcal{A}$ is the amplitude of the waveform
(similar to Ref.~\cite{messenger_read_2012}). We parameterize the tidal piece of the waveform using the
parameterization mentioned in Eq.~(\ref{eq:lambda_m}). We use the above signal model and a Gaussian prior on
$\lambdazero$ with few choices of standard deviation to see how the width in the $\lambdazero$ prior affects
the accuracy in the extraction of the redshift $z$. The injected value of $\lambdazero$ is equal to 200. 
The result for the fractional uncertainty in redshift is shown in the top panel of
Fig.~\ref{fig:h0_uncertainty_gaussian_prior}. We observe a similar trend in the fractional error in
recovery of the redshift as Ref.~\citep{messenger_read_2012} (the $\delta\lambdazero = 0$ case). When
we have a measurement uncertainty, represented by the $\delta\lambdazero = 10, 30$ cases, we find that
it does not affect the measurement uncertainty of the redshift at larger distances. {We
note, however, that the Fisher approximation holds true only in the high signal-to-noise limit. In
the top panel of Fig.~\ref{fig:h0_uncertainty_gaussian_prior} we use a value of SNR $\gtrsim 30$,
shown by the shading, as representing the region where the Fisher approximation is valid.}



To verify these Fisher estimates, we repeat some of the representative, parameter-estimation runs in the CE
era using a Gaussian prior on {\lambdazero}, around the true value of 200, with a standard deviation of 30.
This would correspond to $\sim 15\%$ statistical uncertainty from the value of {$\lambdazero=200$}.
We show the fractional uncertainty in {\hubble, \dl, $z$} in bottom panel of Fig.~\ref{fig:h0_uncertainty_gaussian_prior}.
In this case we mean $90\%$ intervals by $\Delta z, \Delta\dl,\Delta\hubble$.
We observe that the redshift error trends from the full parameter-estimation runs are in agreement with the trends
of the Fisher estimates. In particular, the fractional error in {\hubble} does not change significantly, 
even when we include a Gaussian prior in $\lambdazero$.
This analysis also agrees with the Fisher uncertainties reported in \citeauthor{messenger_read_2012}.
Given these results, we do not expect that a prior statistical uncertainty
in the measurement of {\lambdazero} will significantly affect the final
{\hubble} measurement, especially since most detections will be at larger distances
in the third-generation detector era.

\subsection{Effect of a systematic bias in {\lambdazero}}
\begin{figure}[h!]
    \centering
    \includegraphics[width=0.92\columnwidth, trim=0.1cm 0cm 0cm 1.2cm]
        {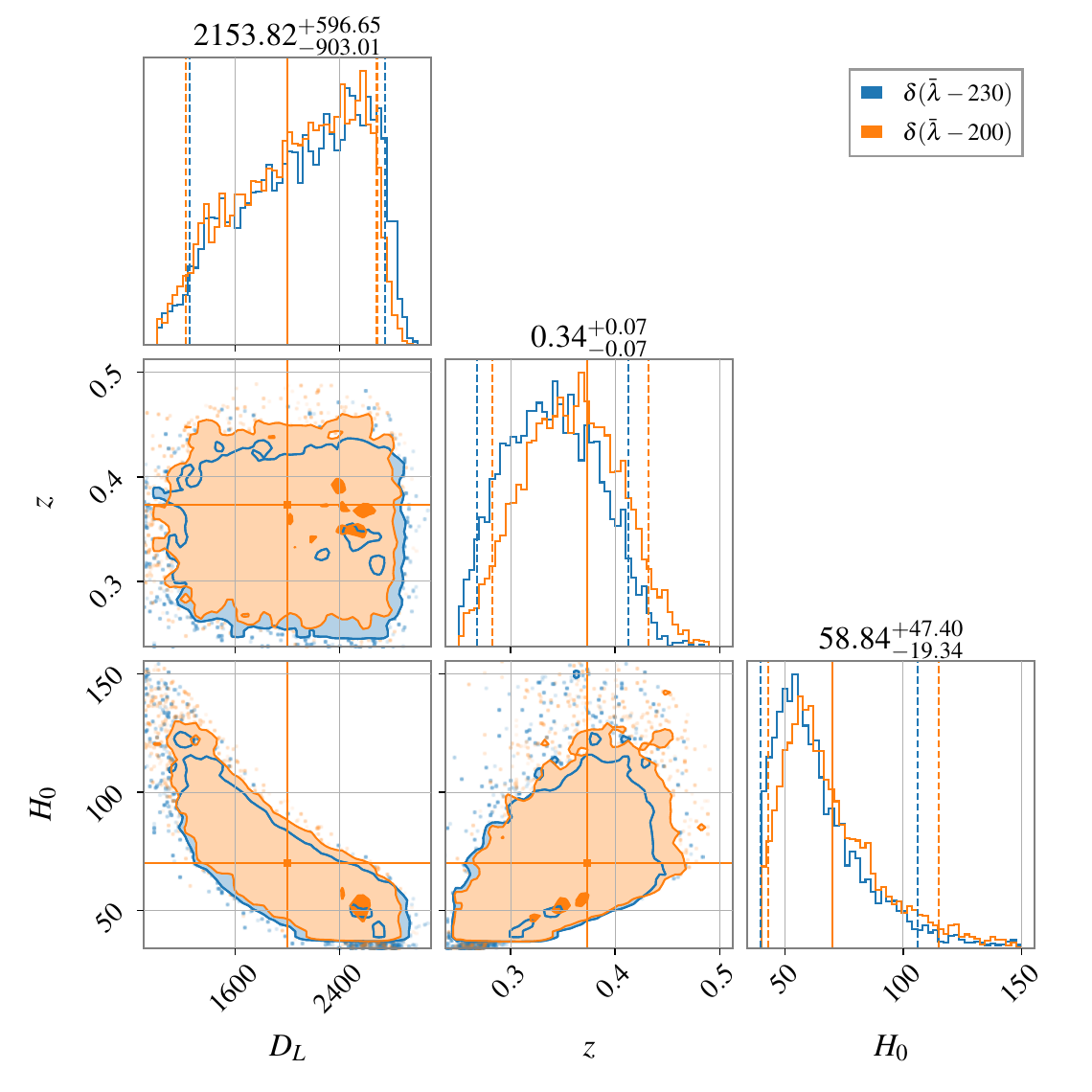}
    \includegraphics[width=0.9\columnwidth, trim=0cm 0.4cm 0cm 1cm, clip]
        {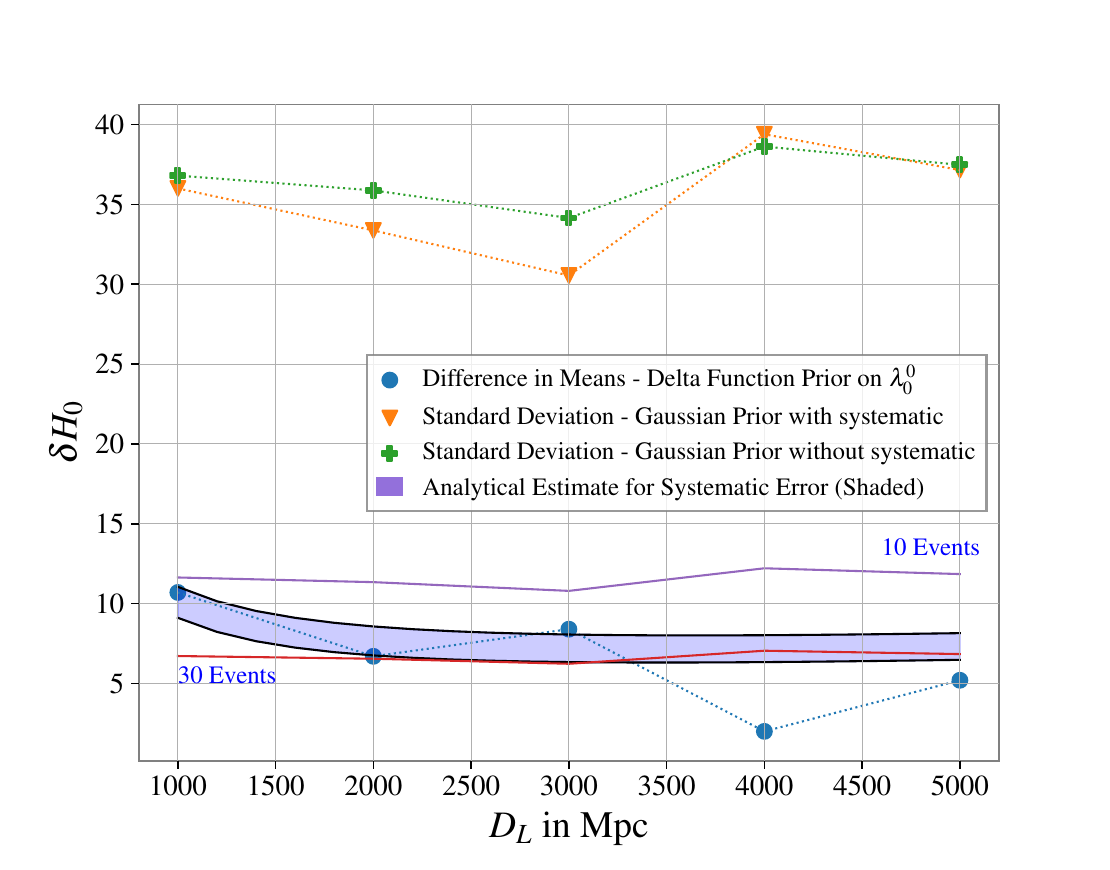}
    \caption{
    \textbf{Top panel}: Comparison of distance, redshift, and {\hubble} measurements for
    an example parameter estimation run injected at a $\dl=2$~Gpc in the CE era.
    Two cases of recovery are shown -- one
    using a prior at the true value $\delta(\lambdazero - 200)$, and another shifted
    $\delta(\lambdazero - 230)$ prior. We observe no significant change in the
    distance recovery, but a systematic shift in the recovery of the redshift;
    the latter impacts the {\hubble} measurement.
    \textbf{Bottom panel}: The difference in means of the {\hubble}
    measurements between the true and shifted cases by doing a few more parameter
    estimation runs as above. This is shown by the dotted line with filled circles.
    We repeat the runs, putting a Gaussian prior with standard deviation of 30 for 
    $\lambdazero$, centered at the true value of 200, and plot half of the 90\% credible
    interval for {\hubble} in dotted line with a plus marker. We repeat one more time
    with the same width but now centered at 230, and plot half credible region in
    using dotted lines with inverted triangles. We find that the credible region
    is not affected by a systematic shift in $\lambdazero$. Therefore, we consider
    the former, and show improvements ($\propto 1/\sqrt{\Ndet}$) as we increase the
    number of detections, using solid lines. Observe that for $\Ndet\sim 30$, the
    statistical improvement hits the boundary of systematic error, shown in blue. We also
    show the analytical estimate for the systematic error from Eq.~(\ref{eq:dz-dlambda-h0})
    using the shading, which roughly agrees with the parameter estimation runs.}
    \label{fig:sys-error-lambda00}
\end{figure}
Following Eq.~(\ref{eq:lambda_h0}), a systematic bias in the measurement of {\lambdazero}
will lead to a biased measurement of $z$, and hence a bias in the inferred value of
{\hubble}. We can estimate this in the following way. Assume first that there is
no bias in the measurement of quantities that depend directly on the signal, like
masses and tidal deformabilities. If so, any systematic bias in {\hubble} will be solely
due to an induced bias in $z$ due to the bias in $\lambdazero$ given the 
parameterization in Eq.~(\ref{eq:lambda_h0}). If one considers then a systematic
bias shift  $\delta\lambdazero$, one finds from Eq.~(\ref{eq:lambda_h0}) that
\begin{equation}
    \delta\lambdatilde = \delta\lambdazero + \delta\left[
        \sum_{k=1}^{3} \frac{\lambdatilde^{(k)}_{0}}{k!}\left( 1 - \frac{\mdet/m_0}{1+z}\right)^k
        \right].
    \label{eq:dz-dlambda-h0}
\end{equation}
A covariance then arises between 
$\delta\hubble$ and $\delta\lambdazero$ that can be explored by setting $\delta \lambdatilde = 0 = \delta\mdet$. The quantity  
$\delta\hubble = (\delta\hubble/\delta\lambdazero) \delta\lambdazero$ is plotted in 
Fig.~\ref{fig:sys-error-lambda00} with $\delta\lambdazero = 30$,
where the upper (lower) limits of the shading correspond to $\msource = 1.3\;(1.5) \msun$.
Observe that the induced error in {\hubble} is much smaller than any statistical error reported earlier. 

In order to verify this estimate, we repeat some of the representative parameter estimation studies,
but this time with models that either have a delta-function or a Gaussian prior that are both peaked
at a shifted location of $\lambdazero = 230$. With these
priors, the redshift measurement shifts from its true values. An example is shown in
the top panel of Fig.~\ref{fig:sys-error-lambda00} where we observe that the recovered redshift is
systematically lower for the case when we use the $\delta(\lambdazero - 230)$ prior. The
distance measurement is unchanged since its information comes from the amplitude which is
not affect due to a biased $\lambdazero$ measurement. The shift in $z$ shows up as a bias
in {\hubble}. We perform parameter estimation for a few more such injections at different distances
with this shifted prior and simply use the difference in means as a benchmark for a systematic error.
In the bottom panel of Fig.~\ref{fig:sys-error-lambda00} we denote this shift in the means
using the dotted line with filled circle markers. We find that it roughly follows the analytic
trend from Eq.~(\ref{eq:dz-dlambda-h0}), shown by the shaded region in the bottom panel of
Fig.~\ref{fig:sys-error-lambda00}.

In reality, we expect the total uncertainty to be more a combination of statistical and systematic,
but as more events are stacked, the statistical error will decrease (as $1/\sqrt{\Ndet}$ or faster),
while the systematic error will not. This then means the latter acts as an \textit{uncertainty floor}
that we must contend with when extracting \hubble. One can then roughly determine how many detections
would be needed to reach this floor. In Fig.~\ref{fig:sys-error-lambda00}, we consider half of the
90\% confidence interval of the {\hubble} measurement for our runs here (shown in dotted lines with
inverted triangle and plus markers) as the statistical uncertainity, and consider an improvement by
$1/\sqrt{\Ndet}$ for $\Ndet = 10, 30$ represented by solid lines in the figure. We see then that
the statistical error will become smaller than the systematic after we have stacked $\Ndet > 30$ events.
However, it is to be noted that this is only illustrated as an example where there is a
systematic bias of 30 from the true value of 200 when measuring $\lambdazero$.


\subsection{Effect of uncertainty in binary-love relations}\label{subsec:systematics_binary_love}
\begin{figure}
    \centering
    \includegraphics[width=0.5\textwidth]{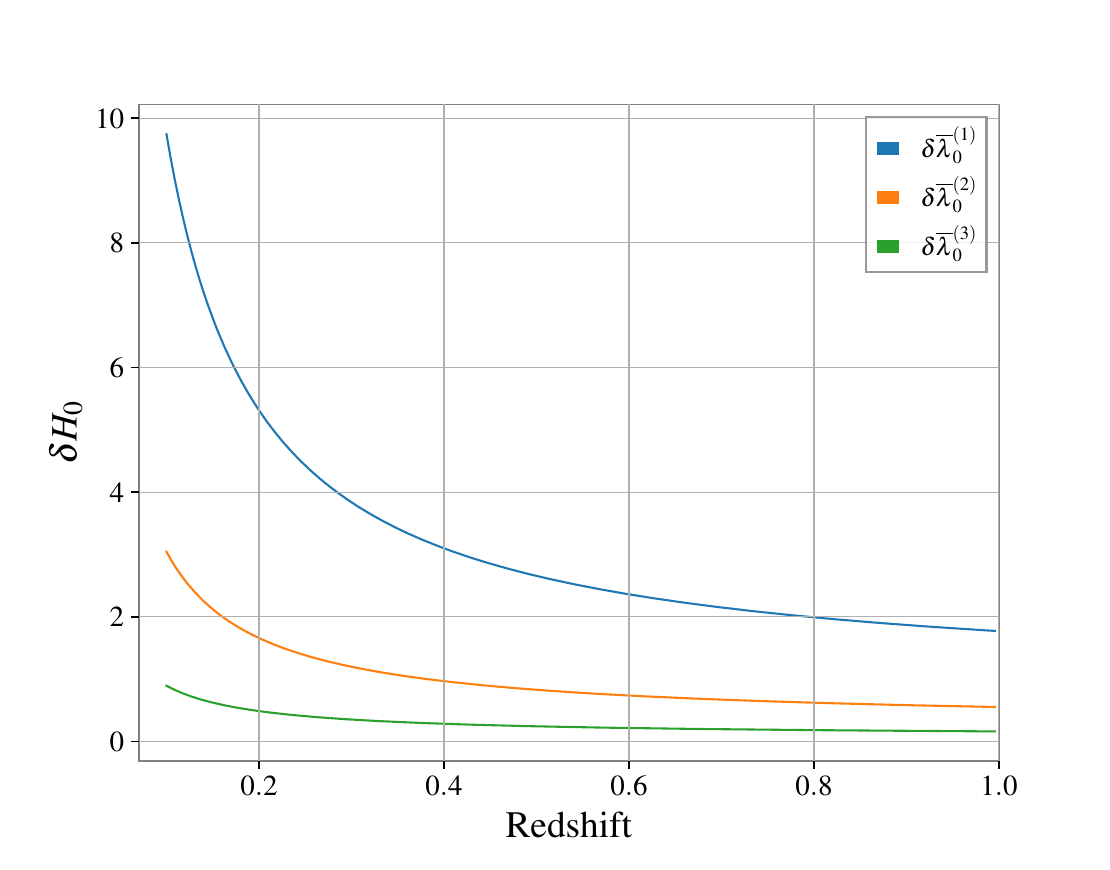}
    \caption{An analytic estimate of the bias in {\hubble} caused by assuming the non-universality of the
    binary-love relations (see Sec.~\ref{subsec:systematics_binary_love}). As an example, here we consider
    the true EoS to be MPA1 which has the largest residual among the {\numeos} EoSs from the fit in Eq.~(\ref{eq:lambda_k_relativistic})
    We see that the error due to loss of universality reduces for $\lambdak$ as we increase $k$.}
    \label{fig:sys-error-EOS}
\end{figure}
We end this section with an estimate of the systematic error in the inference of {\hubble} induced by our assumption that the binary love relations are exact and EoS independent. In order to arrive at a rough estimate, we consider how a bias in $\lambdak$, given a value of $\lambdazero$, would affect our measurements of \hubble. In order to separate this effect from those discussed in the previous subsections, we assume now that $\lambdazero$ has been measured perfectly. With this in mind, the bias in the $\lambdak$ for $k\geq1$ inside the summation in
Eq.~(\ref{eq:dz-dlambda-h0}) leads to a bias in the inferred redshift, and hence a bias in {\hubble}. This can be estimated by taking a variation of the individual terms as,
\begin{equation}
    \delta\left[\frac{\lambdatilde^{(k)}_{0}}{k!}\left( 1 - \frac{\mdet/m_0}{1+z}\right)^k
    \right] = 0.
\end{equation}
This expression can be solved for $\delta\hubble/\delta\lambdak$ and multiplied with $\delta\lambdak$ 
to obtain the bias in {\hubble}. This is shown in Fig.~\ref{fig:sys-error-EOS}.
Here, we choose the residuals for the MPA1 EoS as the values for $\delta\lambdak;\;k \geq 1$.
Observe that the systematic error introduced in the extraction of {\hubble} is much smaller than the statistical error.
From this we conclude that only after more than a certain number
of events are stacked ($\sim 30$ in this case), will this systematic error be of any importance.
Observe also that the systematic error due to the binary Love relations is smaller for the higher-order terms in the Taylor expansion. 

\section{Conclusion}\label{sec:conclusion}
This work demonstrates yet another application of universal relations in NSs. Following
the proposition by~\citet{messenger_read_2012}, previous attempts at measuring the distance-redshift 
relation solely using GWs have relied on the knowledge of the specific NS EoS.~\footnote{
For other techniques that use population properties see
Refs.~\cite{Taylor_2012,Mukherjee_2021,mastrogiovanni2021cosmology}}
The expansion
of the tidal deformability in terms of the source mass is particularly interesting since it breaks
the degeneracy between GW frequency and redshift. However, in absence of \emph{a priori} EoS information,
all of the expansion coefficients are free parameters. The $\lambdazero - \lambdak$ relations
greatly constrain this degree of freedom. By employing these relations, knowing only a single
coefficient determines the rest. This can be
particularly useful for setting physically motivated priors on the tidal deformability. In
GW astrophysics, previous literature has considered putting flat priors on $\lambdatilde_{1,2}$
or $\bar{\Lambda}$. However, another option which is physically motivated prior is an uninformative
prior on the free universal expansion coefficient, $\lambdazero$, which in turn restricts the
priors on $\lambdatilde_{1,2}$, or $\bar{\Lambda}$. In this work, we employ this
technique to GW170817 strain data, and obtain the measurement of $\lambdatilde(1.4\msun)$ that
is consistent with previous measurements by the LIGO/Virgo collaboration. 

The advantage of using the $\lambdazero - \lambdak$ relations is that the tidal deformability
is only parameterized by two quantities: $\lambdazero$ and $\msource$ (or equivalently, $z$).
Future multi-messenger observations of GWs from BNSs with simultaneous identification of the
redshift would lead to measurements of $\lambdazero$ that can be combined to give a constrained
measurement for this fundamental quantity. Such observations are well motivated based on
forecasting studies of different observing scenarios of GW observations~\cite{observing_scenarios}
combined with development of current and next generation synoptic surveys, and cyber-infrastructure
developments.~\footnote{For example, SCiMMA~(\url{https://scimma.org/}).}
With a constraint on $\lambdazero$, the use case can be flipped to now measure the redshift of
BNS signals from the tidal deformability measurements.
This is advantageous since the method does not rely on any prompt follow-up operations, and is
also free from selection effect of host galaxy identification or galaxy catalog incompleteness
that has been used in the literature till date.

We demonstrate this technique of measuring the Hubble constant, \hubble, using a synthetic
population of detected NS binaries.  We also analyze the impact of error in measurement of
{\hubble} due to the statistical and systematic errors in this prescription.
However, aside from GW observations constraining allowed NS EoSs, there are other missions
which are also putting stringent measurement of the mass and radii of NSs.
Recently, the NICER team reported the discovery
of X-ray pulses from massive milli-second pulsars, PSR J0740+6620 and PSR J1614-2230~\cite{2021arXiv210506978W}.
The data from PSR J0740+6620 has been used to measure the mass-radius relation of
the NS~\cite{2021arXiv210506979M, 2021arXiv210506980R, 2021arXiv210506981R}.
Such measurements are expected to rule out representative EoSs from the literature
that are inconsistent with observables, further reducing the uncertainty in the
$\lambdazero - \lambdak$ relations presented here.

\begin{acknowledgements}
D.~C. is supported by the Illinois Survey Science Fellowship
from the Center for AstroPhysical Surveys (CAPS)~\footnote{\url{https://caps.ncsa.illinois.edu/}}
at the National Center for Supercomputing Applications (NCSA), University of Illinois Urbana-Champaign.
D.~C. acknowledges computing resources provided by CAPS to carry our this research.
K.Y. acknowledges support from NSF Grant PHY-1806776, NASA Grant 80NSSC20K0523, a Sloan Foundation Research Fellowship and the Owens Family Foundation. 
K.Y. would like to also acknowledge support by the COST Action GWverse CA16104 and JSPS KAKENHI Grants No. JP17H06358.
G.~H.~, D.~.E.~H.~, A.~H.~and N.~Y.~ acknowledge support from NSF grant AST Award No.~2009268.
This work made use of the Illinois Campus Cluster, a computing resource that is operated by the Illinois Campus
Cluster Program (ICCP) in conjunction with NCSA, and is supported by funds from the University of Illinois at
Urbana-Champaign.

This research has also made use of data obtained from the Gravitational Wave Open Science
Center (\url{https://www.gw-openscience.org/}),~\cite{Rich_Abbott_2021} a service of LIGO Laboratory, the LIGO Scientific
Collaboration and the Virgo Collaboration. LIGO Laboratory and Advanced LIGO are funded by the United States National Science
Foundation (NSF) as well as the Science and Technology Facilities Council (STFC) of the United Kingdom, the Max-Planck-Society
(MPS), and the State of Niedersachsen/Germany for support of the construction of Advanced LIGO and construction and operation
of the GEO600 detector. Additional support for Advanced LIGO was provided by the Australian Research Council. Virgo is
funded, through the European Gravitational Observatory (EGO), by the French Centre National de Recherche Scientifique
(CNRS), the Italian Istituto Nazionale di Fisica Nucleare (INFN) and the Dutch Nikhef, with contributions by institutions
from Belgium, Germany, Greece, Hungary, Ireland, Japan, Monaco, Poland, Portugal, Spain.

The authors would like to thank Jocelyn Read for reviewing the document and providing helpful feedback.
This document is given the LIGO DCC number P2100195.~\footnote{\tiny{\url{https://dcc.ligo.org/LIGO-P2100195}}}
The authors would also like to thank the anonymous referee for helpful comments.
\end{acknowledgements}

\newpage
\appendix
\section{Implied prior on $\lambdatilde_{1,2}$ by $\lambdazero$}\label{appendix:lambda_1_2_prior}
\begin{figure}
    \centering
    \includegraphics[width=1.0\columnwidth]{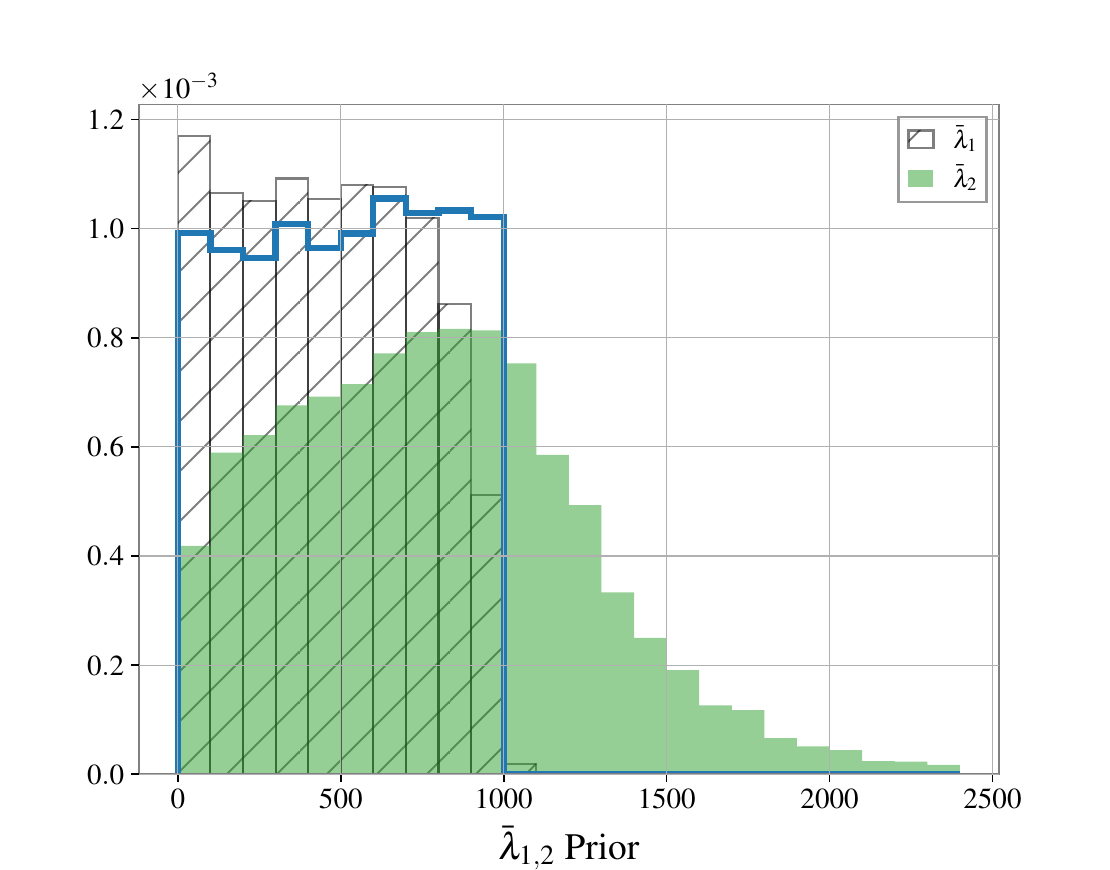}
    \caption{The implied (marginal) priors on the individual tidal deformability
    parameters due to the uniform prior in $\lambdazero$ in Sec.~\ref{sec:pe}.
    The prior on $\lambdatilde_1$ is shown in the hatched histogram, and
    $\lambdatilde_2$ is shown in the filled histogram. The $\lambdazero$
    prior is shown in the unfilled histogram.}
    \label{fig:lambda12_prior}
\end{figure}
\begin{figure*}
    \centering
    \includegraphics[width=0.48\textwidth]{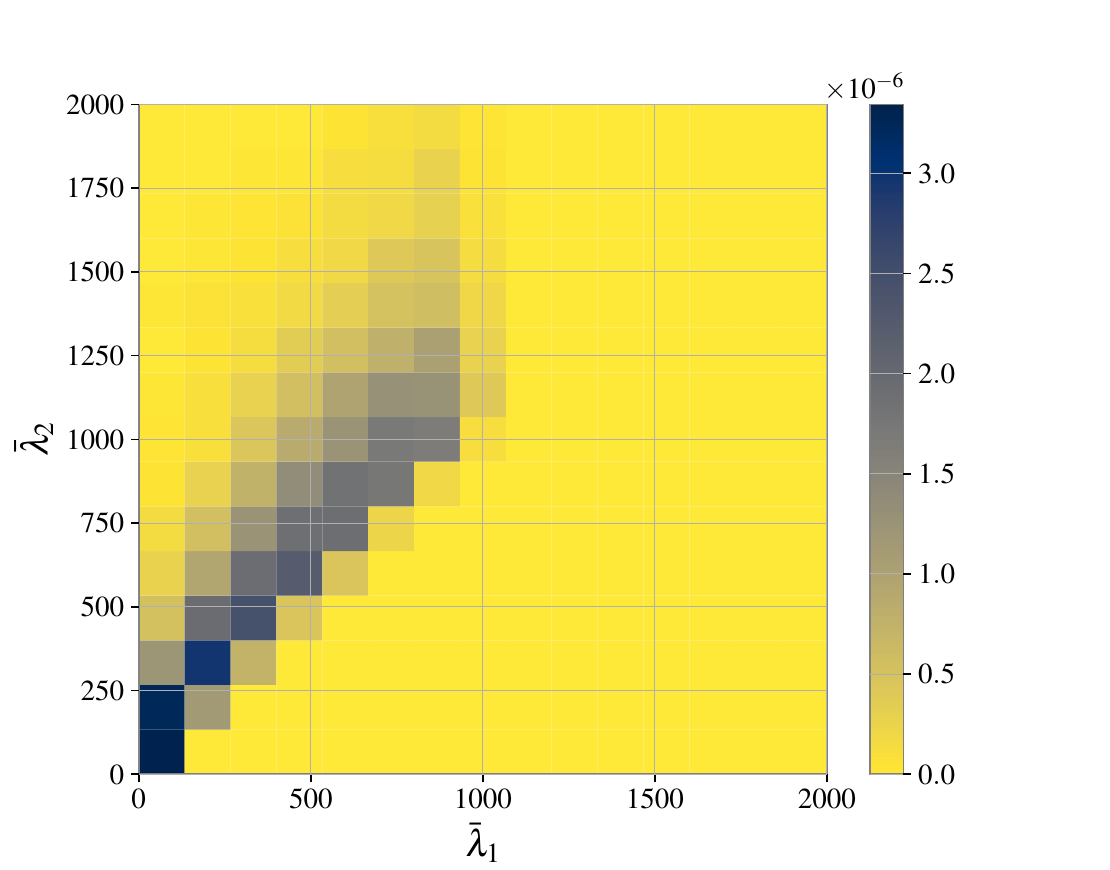}
    \includegraphics[width=0.48\textwidth]{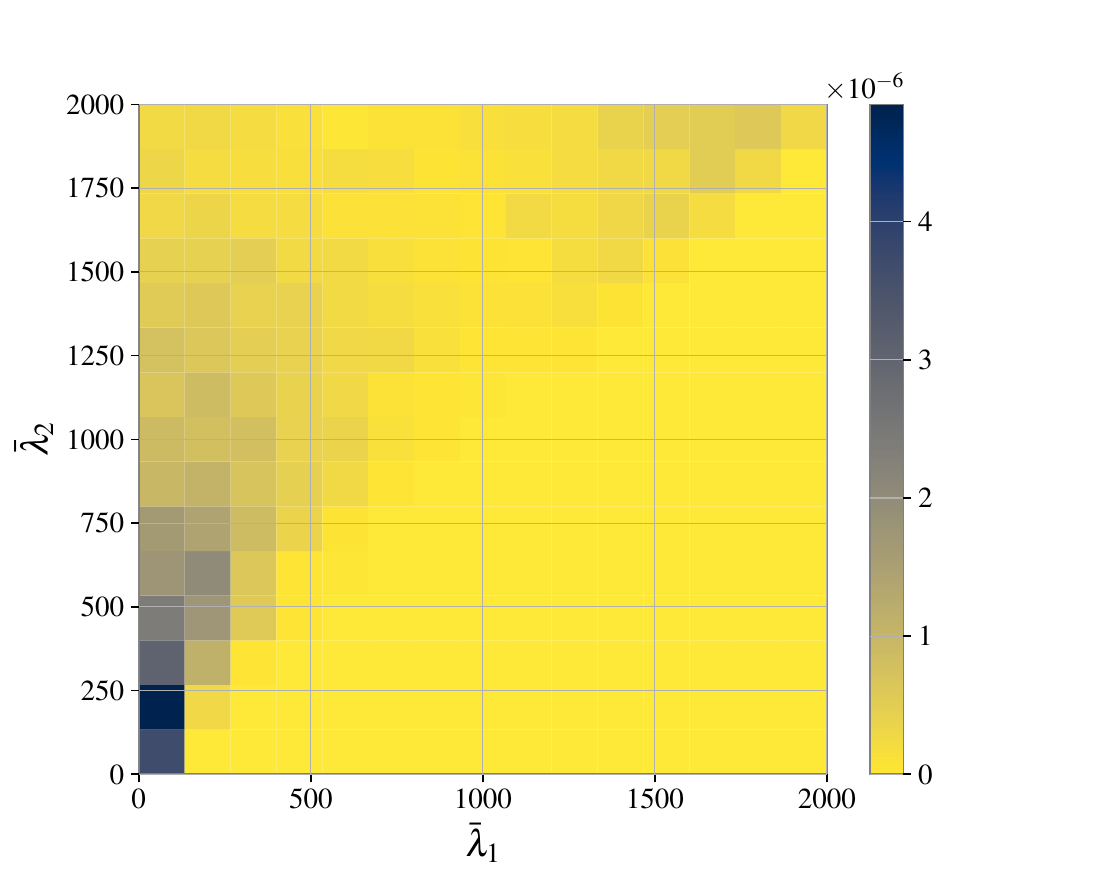}
    \caption{\textbf{Left panel}: Non-marginalized version of Fig.~\ref{fig:lambda12_prior}. Here,
    we have used the parameterization using $\lambdazero-\lambdak$ relations from Sec.~\ref{sec:lambda_0}.
    Observe the correlation between $\lambdatilde_{1,2}$, which is expected due to the prior
    information about the masses when imposing a common EoS.
    \textbf{Right panel}: The prior on implied  $\lambdatilde_{1,2}$ when using the
    universal relation between the symmetric and anti-symmetric tidal deformability,
    $\lambdatilde_s - \lambdatilde_a$, relations from YY17. This was used in the EoS-insensitive
    results reported previously in Ref.~\cite{2018PhRvL.121p1101A}.
    }
    \label{fig:appendix_lambda_1_lambda_2_prior}
\end{figure*}
In Sec.~\ref{sec:pe}, we did not set an explicit prior on the individual tidal
deformabilities, $\lambdatilde_{1,2}$ independently. However, the uniform prior
on the $\lambdazero$ parameter along with the prior on the individual component
masses imply a prior on the individual tidal deformabilities, $\lambdatilde_{1,2}$.
We show this in Fig.~\ref{fig:lambda12_prior}. It should be noted that there are
correlations between the two tidal deformabilities based on the priors on the masses.
We show this in Fig.~\ref{fig:appendix_lambda_1_lambda_2_prior}. On the left panel therein,
we see the correlation between the $\lambdatilde_{1,2}$. Here, for the mass priors, we
use the same mass priors as Sec.~\ref{sec:pe}. The correlation is motivated from
the fact that in the limit of equal mass components, we expect the tidal deformabilities
to be equal. We also note that the prior implied in this work differs from that used in
Ref.~\cite{2018PhRvL.121p1101A} where instead the symmetric tidal deformability,
$\lambdatilde_s = (\lambdatilde_1 + \lambdatilde_2)/2$ was sampled uniformly, and then
$\lambdatilde_a = (\lambdatilde_1 - \lambdatilde_2)/2$ was obtained using another
universal relation, $\lambdatilde_a = \lambdatilde_a(\lambdatilde_s, q)$, reported
in YY17. We reproduce the implied prior obtained using the latter technique in the right
panel of Fig.~\ref{fig:appendix_lambda_1_lambda_2_prior}. We attributed the differences
in the upper limit value for $\lambdazero$ between this work and Ref.~\cite{2018PhRvL.121p1101A},
mentioned in Sec.~\ref{sec:pe} due to the difference in priors.

\section{Parameter estimation from GW170817 data}\label{appendix:pe}
\begin{figure*}
    \centering
    \includegraphics[width=0.9\textwidth]{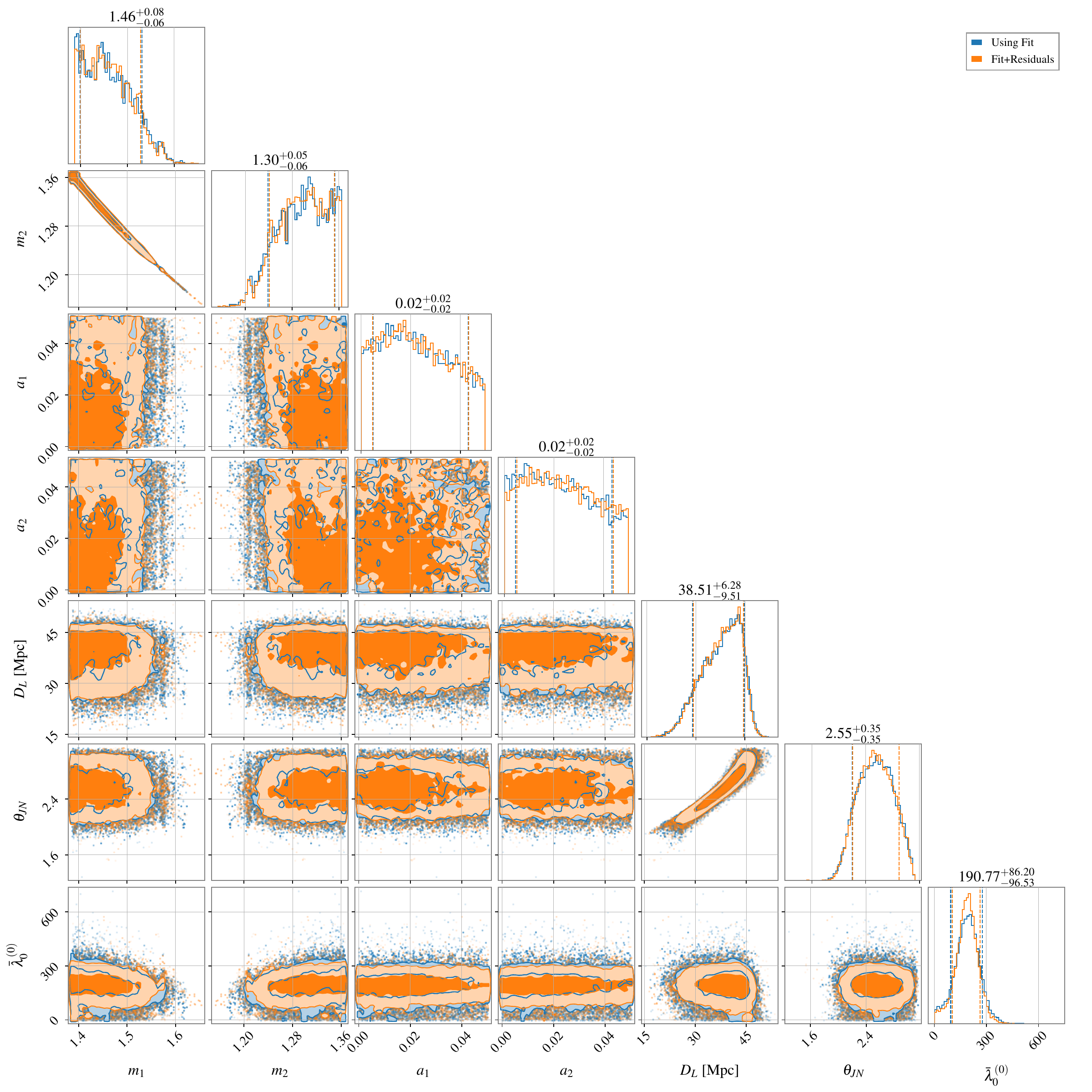}
    \caption{Extended (corner) plot showing pair and marginal parameters from performing Bayesian
    parameter estimation on GW170817 data discussed in Sec.~\ref{sec:pe}. The
    parameterization of the tidal deformabilities follows Eq.~(\ref{eq:lambda_m}).
    The redshift of the source is fixed to 0.0099. The bottom right posterior
    for {\lambdazero} is the same as Fig.~\ref{fig:lambda_zero_posterior} in the
    main text. We show the distribution of parameters in both cases - when using
    the fit in Eq.~(\ref{eq:lambda_m}) directly, and when including the residual
    errors from the fit mentioned in the Sec.~\ref{sec:lambda_0}.}
    \label{fig:corner-plot}
\end{figure*}
In Sec.~\ref{sec:lambda_0}, we showed the marginalized posterior on the
universal quantity, \lambdazero, from performing Bayesian parameter estimation.
Here, we show pair-plots and marginalized distributions (corner plot) of some of
the other parameters in Fig.~\ref{fig:corner-plot}. The priors on the masses are
the marginalized posterior of the detector-frame values reported in GWTC-1. For
the luminosity distance, we use a uniform in comoving volume up to 75~Mpc.
For the spins, we use the low-spin prior from GWTC-1. The bottom right block
is the same as Fig.~\ref{fig:lambda_zero_posterior} in the main body of the text.
For the {\bilby} configuration, we use the nested sampling algorithm, {\dynesty},
with number of live points \texttt{nlive = 1500}, and number of autocorrelation
lengths to reject before accepting a new point, \texttt{nact = 10}. These settings
are motivated from settings used in the validation of bilby against GWTC-1
events~\cite{Romero_Shaw_2020}.

\section{Trends in {\hubble} recovery with inclination}\label{appendix:hubble-trends}
\begin{figure}[h!]
    \centering
    \includegraphics[width=1.0\columnwidth, trim=1cm 1cm 1cm 0cm]{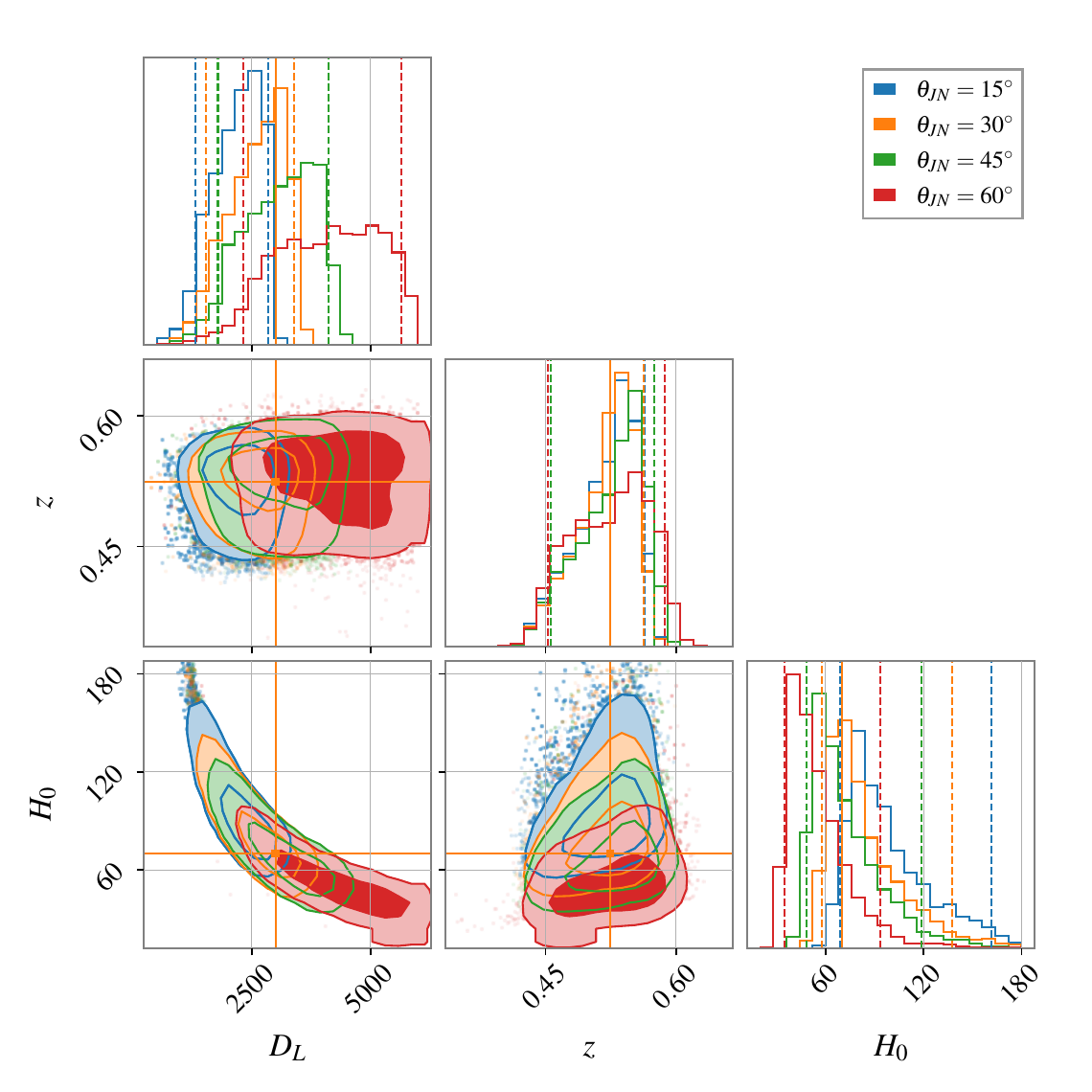}
    \caption{This is an example of {\dl}, $z$, and {\hubble} measurements for
    different combinations of the true inclination angle, $\theta_{JN}$, at a fixed
    distance of 3~Gpc in the Cosmic Explorer era. Note that with increasing true inclination
    angle, there is increasing support for larger distance values. This is expected since a larger
    inclination implies a lower injected amplitude, which is recovered as a larger distance.
    The redshift measurement is not affected since it comes from the phase of the waveform.
    }
    \label{fig:appendix_example_z_hubble_distance}
\end{figure}
In this section we illustrate how the inclination angle affect the distance
recovery, which impacts the posterior of {\hubble}.
In Fig.~\ref{fig:appendix_example_z_hubble_distance}, we show the distance, redshift,
and {\hubble} recovery considering injections at a fixed distance but varying
inclination angles. We observe that the recovered distance has support for larger
values as the true inclination angle increases. This is expected since edge-on
sources have lower SNR compared to face-on sources, and therefore are degenerate with
a larger distance recovery. On the other hand, since the redshift is recovered from the
phase of the waveform, it is not affected. This implies that the {\hubble} recovery
is affected in the opposite sense, having support for higher-than-true values for
face-on sources and vice versa. However, in all cases, there is support for the
true value of $70~\hubbleunit$.

\section{Considering alternative redshift priors}\label{appendix:alternative_prior}
\begin{figure}
    \centering
    \includegraphics[width=1.0\columnwidth]
        {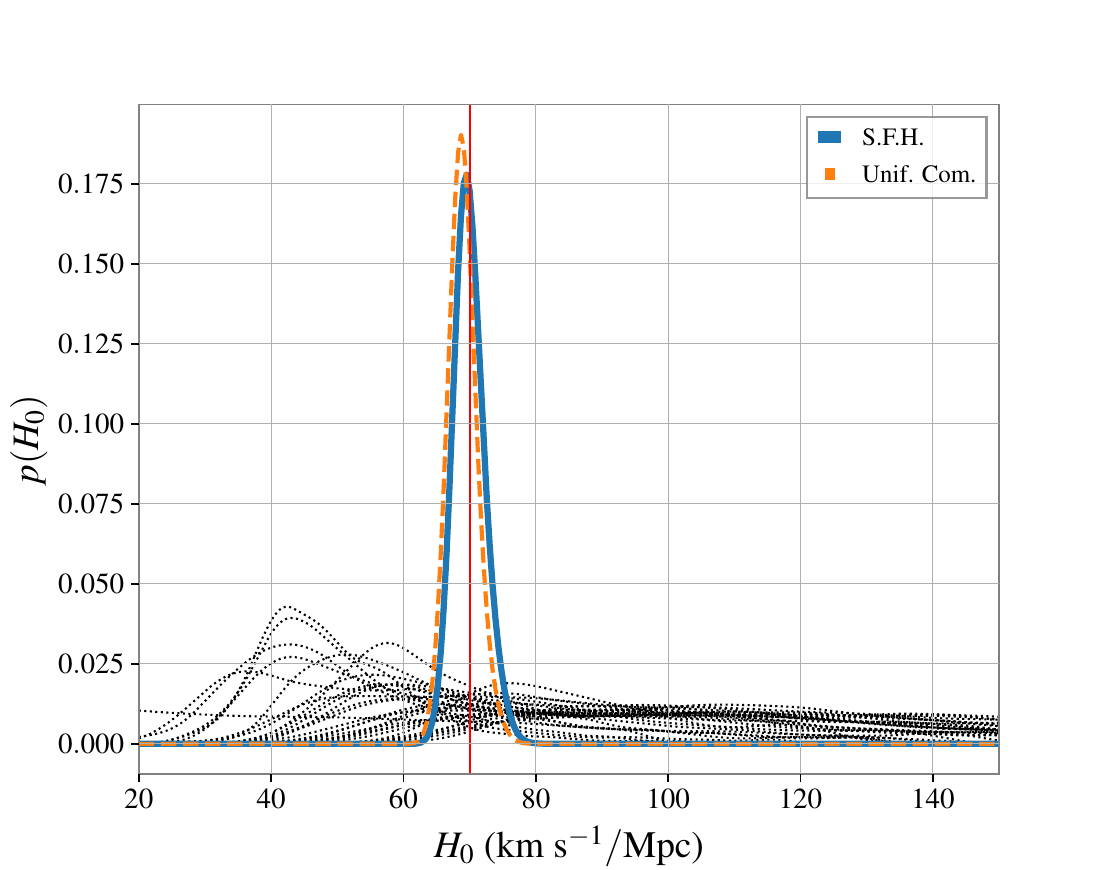} \\
    \includegraphics[width=1.0\columnwidth, trim=0cm 0cm 1cm 0cm, clip]
        {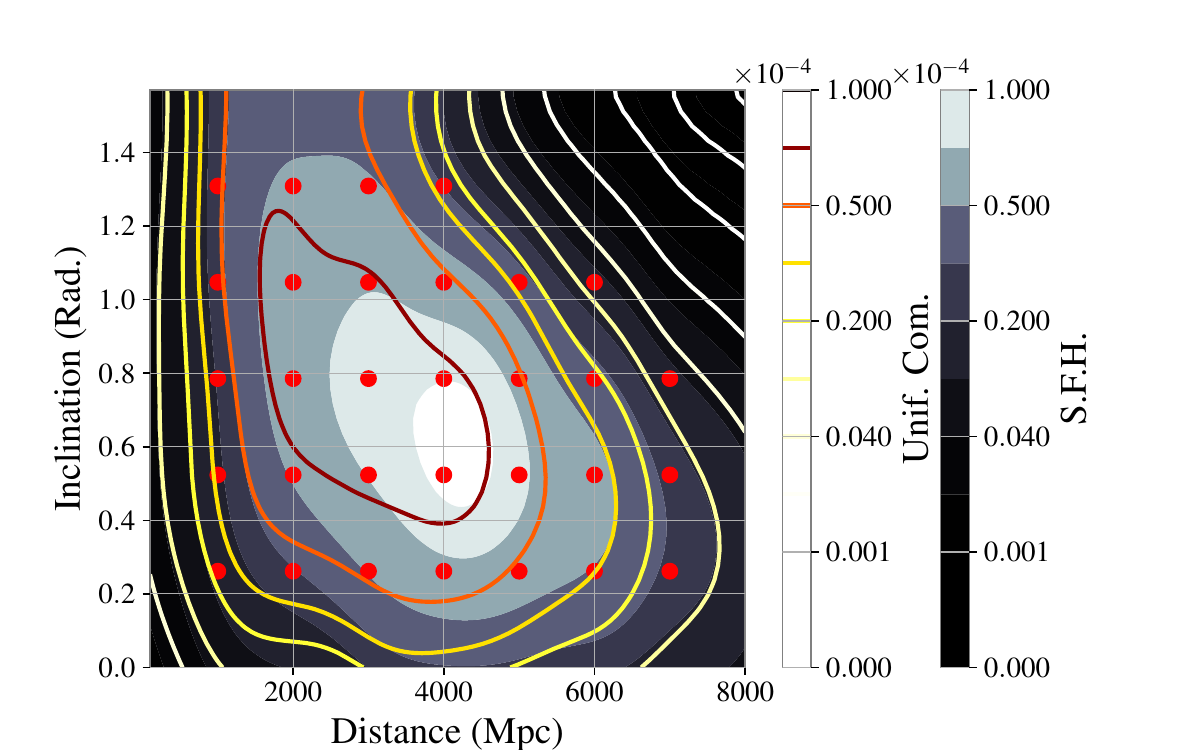}
    \caption{\textbf{Upper panel}: The combined measurement of the {\hubble},
    similar to Fig.~\ref{fig:stacked_h0} for the Cosmic Explorer example.
    Here, we have reweighted the normalized number count of the recovered
    events using a redshift prior following uniform in comoving volume (dashed), and
    that following star formation history (solid).
    \textbf{Bottom panel}: The contour plot showing the normalized number counts
    similar to Fig.~\ref{fig:stacked_h0} for the two alternative priors. We observe
    that the combined measurement of {\hubble} is not significantly affected
    due to the prior on redshift.
    }
    \label{fig:appendix_other_distr}
\end{figure}
In this appendix, we show the effect of using alternative choices for
distances/redshift compared to distributing them uniformly in volume, $\propto \dl^2$,
used in the main body of the text for Fig.~\ref{fig:stacked_h0}. We consider the Cosmic
Explorer example from Sec.~\ref{sec:hubble}. We consider two cases -- 1) redshift
distribution such that the rate is uniform in comoving volume, and 2) it follows
the cosmic star formation history~\citep[following Eq.~15 from Ref.][]{Madau_2014}.
We re-weight the luminosity distance of recovered binary systems in Sec.~\ref{sec:hubble}
using the two priors as shown in Fig.~\ref{fig:appendix_other_distr}. For Bayesian
parameter estimation, we consider the same representative distance/inclination gridpoints
as in the Cosmic Explorer panel of Fig.~\ref{fig:stacked_h0}, but count them based on the
re-weighted heatmap in Fig.~\ref{fig:appendix_other_distr}. We find that the combined
{\hubble} measurement is not affected by the choice of the priors. We also would like
to note that while the evolution of the BBH rate has been done~\cite{pop_dist_gwtc_2},
the BNS rate has not been constrained strongly due to lack of BNS observations 
and their relative low distances compared to BBHs. Hence, we feel that our choice
made here is justified. The analysis can be redone as further constrains are put on the same.

\section{Reweighting the {\hubble} posterior}\label{appendix:reweighting}
\begin{figure}
    \centering
    \includegraphics[width=1.0\columnwidth]{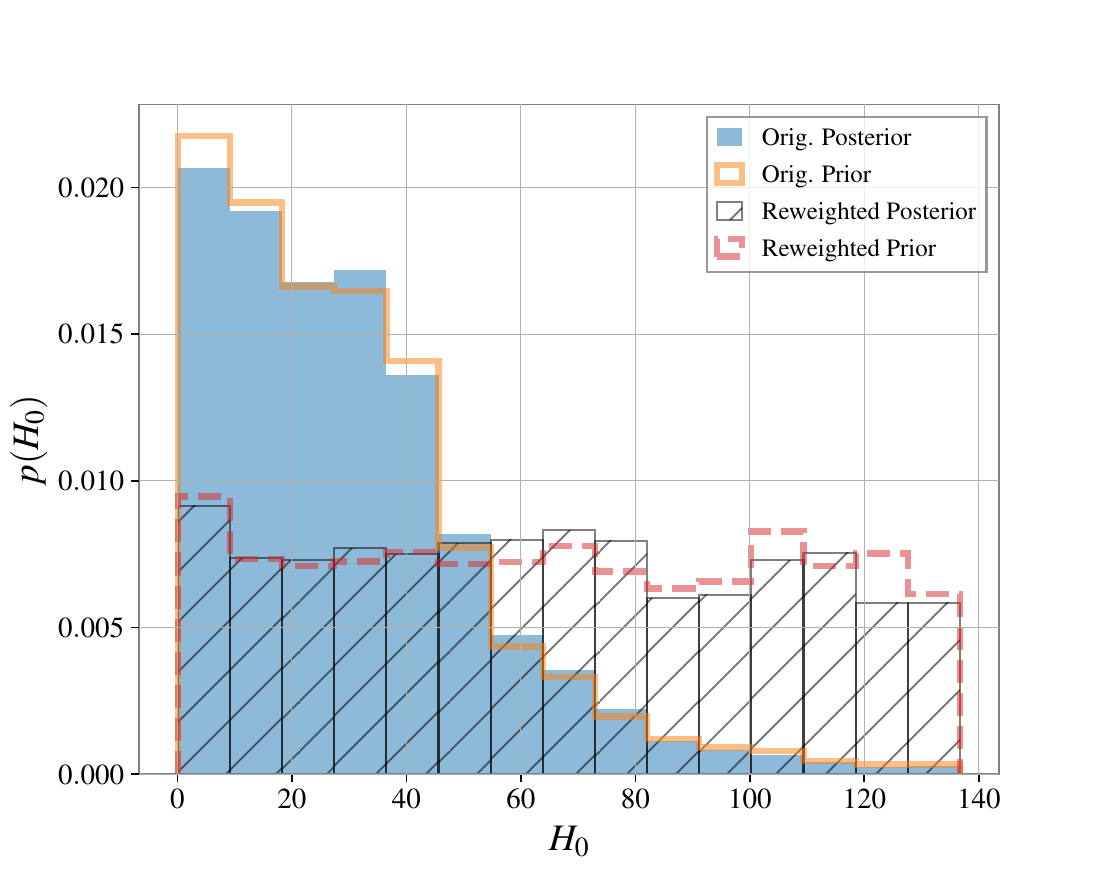}
    \caption{Illustration of obtaining the likelihood from the {\hubble} posterior.
    In this example, we injected a low SNR ($\sim 1$) injection, and perform parameter
    estimation mentioned in Appendix~\ref{appendix:pbilby}. The original prior on {\hubble}
    is shown by solid unfilled histogram, the posterior is shown in filled histogram.
    The re-weighted prior and posterior are shown as dashed and hatched histograms. The
    latter is what is expected in absence of any information.}
    \label{fig:appendx_reweighting}
\end{figure}
In Sec.~\ref{sec:hubble}, we employ the $\lambdazero - \lambdak$ relation, to
parameterize the tidal deformability in terms of the redshift $z$. Therefore,
one can either sample in $z$, or with the other cosmological parameters fixed in a
{\lcdm} universe, sample directly in the {\hubble}. In the former case, putting a prior
on $z$ and the luminosity distance, $\dl$, implies a prior
on \hubble, while in the latter, we directly put a prior on the {\hubble}. 
In either case, when combining multiple observations, we need to multiply the likelihoods,
dividing out by any imposed or implied prior. In practice putting a uniform prior when sampling
(ensuring that the posterior does not rail against the prior boundaries) is equivalent
of the likelihood up to a constant factor. We illustrate this using a low signal to noise
ratio (SNR $\sim$ 1) event. We put a prior on {\dl} that is uniform in comoving
volume up to 5~Gpc, and a uniform prior on redshift $\in [0, 0.5]$. The implied prior on
{\hubble} due to this prior choice is shown in Fig.~\ref{fig:appendx_reweighting} by the
solid, unfilled histogram. Due to the low SNR of the injection in this case, we expect the
posterior to be similar to the prior on {\hubble}. To obtain the likelihood, we need to divide
out by the prior, or reweight the samples such that the new prior on {\hubble} is flat.
In practice this is done by binning the posterior samples and weighting them by the inverse
count of the prior distribution. The new re-weighted posterior
is shown by the hatched histogram. As expected, this measurement is uninformative i.e., the
re-weighted posterior, which in this case is the likelihood, is flat.

\section{\pbilby~configuration for BNS simulations}\label{appendix:pbilby}
For Sec.~\ref{sec:hubble}, we
make use of the {\pbilby} framework~\citep{Smith_2020} which is an extension of
{\bilby} to scale out the analysis using Message Passing Interface
(MPI)~\footnote{\url{https://mpi4py.readthedocs.io/}} 
to an entire high-performance cluster. We use a fiducial binary with a
source-frame chirp mass of $\mchirp = 1.17~\msun$ and mass-ratio, $q = 0.9$ for all of our
synthetic injections. The signal duration is 128s. While BNS signals will last much
longer during the CE era, the 5PN tidal terms only become pronounced close to merger.
The injected redshifted chirp mass is
obtained as, $(1 + z^{\text{inj}})\mchirp^{\text{inj}}$, where $z^{\text{inj}}$ is
determined from the injected luminosity distance that assumed true
flat-$\lcdm(\hubble=70~\hubbleunit, \Omega_{m0}=0.3)$ cosmology.
For sampling, we use a prior that is uniform in detector-frame
chirp mass between $(1 + z^{\text{inj}})\mchirp^{\text{inj}} \pm 0.1~\msun$,
and a prior uniform in mass-ratio $\in [0.65, 1]$.
We ignore spins (set them to zero), and use delta function priors at zero
on spin when sampling. We also fix the sky-location of the source. The prior for
distance is uniform in comoving volume up to twice the injected distance.
We use a stationary Gaussian noise realization for each of
our detector era. For the O5 and CE results in Fig.~\ref{fig:stacked_h0},
we impose a uniform prior on the redshift, and get the individual likelihoods
using the technique mentioned in Appendix~\ref{appendix:reweighting}. For the Voyager
results, we sample directly in {\hubble} using a uniform prior
$\in [1, 300]~\hubbleunit$.

\newpage
\bibliography{references}
\bibliographystyle{apsrev}
\end{document}